\documentclass{aa}
\usepackage{natbib}
\usepackage{graphicx}
\usepackage{hyperref}
\hypersetup{
    colorlinks=false, 
    linkcolor=blue, 
    urlcolor=red, 
    linktoc=all 
}
\bibpunct{(}{)}{;}{a}{}{,}
\graphicspath{{images/}}
\usepackage{multirow}
\usepackage{comment}
\usepackage{color}
\usepackage{txfonts}
\usepackage{ulem}
\newcommand{\niv}{N\,\textsc{iv}]}
\newcommand{\nv}{N\,\textsc{v}}
\newcommand{\ovi}{O\,\textsc{vi}}
\newcommand{\Niv}{N\,\textsc{iv}]\,$1483, 1486\AA$}
\newcommand{\Nv}{N\,\textsc{v}\,$1240\AA$}
\newcommand{\Ovi}{O\,\textsc{vi}\,$1032\AA$}
\newcommand{\cii}{[C\,\textsc{ii}]}
\newcommand{\Cii}{[C\,\textsc{ii}]$\,158\mu\rm{m}$}
\newcommand{\Lyalpha}{Ly$\alpha$}
\newcommand{\Ciistar}{C\,\textsc{ii}$^*$\,$1335\AA$}
\newcommand{\ciistar}{C\,\textsc{ii}$^*$}
\newcommand{\Civ}{C\,\textsc{iv}\,$1550\AA$}
\newcommand{\Heii}{He\,\textsc{ii}\,$1640\AA$}
\newcommand{\Ciii}{C\,\textsc{iii}]\,$1908\AA$}
\newcommand{\oiii}{[O\,\textsc{iii}]}
\newcommand{\Siiv}{Si\,\textsc{iv}$\,1394, 1403\AA$}
\newcommand{\ebv}{$\rm{E}(\rm{B}-\rm{V})$}

\begin{document}

   \title{The ALPINE-ALMA [CII] survey: double stellar population and AGN activity in a galaxy at $z\sim5.5$}

   \author{L. Barchiesi\thanks{\email{luigi.barchiesi2@unibo.it}}
          \inst{1,2}
          \and
          M. Dessauges-Zavadsky \inst{3}
          \and
          C. Vignali \inst{1,2}
          \and
          F. Pozzi \inst{1,2}
          \and
          R. Marques-Chaves \inst{3}
          \and
          A. Feltre \inst{4,2}
          \and
          A. Faisst \inst{5}
          \and
          M. B\'ethermin \inst{6}
          \and
          P. Cassata \inst{7,8}
          \and
          S. Charlot \inst{9}
          \and
          Y. Fudamoto \inst{10,11}
          \and
          M. Ginolfi \inst{12}
          \and
          E. Ibar \inst{13}
          \and
          G. C. Jones \inst{14}
          \and
          M. Romano \inst{15}
          \and
          D. Schaerer \inst{3}
          \and
          L. Vallini \inst{16}
          \and
          E. Vanzella \inst{2}
          \and
          L. Yan \inst{17}
          }

   \institute{Dipartimento di Fisica e Astronomia, Universit\`a degli Studi di Bologna, via P. Gobetti 93/2, 40129 Bologna, Italy
  \and INAF-OAS, Osservatorio di Astrofisica e Scienza dello Spazio di Bologna, via Gobetti 93/3, 40129 Bologna, Italy
  \and Department of Astronomy, University of Geneva, Chemin Pegasi 51, 1290 Versoix, Switzerland
  \and SISSA, Via Bonomea 265, 34136, Trieste, Italy
  \and IPAC, California Institute of Technology 1200 E California Boulevard, Pasadena, CA 91125, USA
  \and Aix Marseille Univ., CNRS, LAM, Marseille, France
  \and Dipartimento di Fisica e Astronomia, Università di Padova, Vicolo dell’Osservatorio 3, I-35122, Padova, Italy
  \and INAF – Osservatorio Astronomico di Padova, Vicolo dell’Osservatorio 5, I-35122, Padova, Italy
  \and Institut d'Astrophysique de Paris, Sorbonne Universit\'e, CNRS, UMR7095, F-75014 Paris, France
  \and Waseda Research Institute for Science and Engineering, Faculty of Science and Engineering, Waseda University, 3-4-1 Okubo, Shinjuku, Tokyo 169-8555, Japan
  \and National Astronomical Observatory of Japan, 2-21-1, Osawa, Mitaka, Tokyo, Japan
  \and European Southern Observatory, Karl-Schwarzschild-Strasse 2, 85748, Garching, Germany
  \and Instituto de F\'isica y Astronom\'ia, Universidad de Valpara\'iso, Avda. Gran Breta\~na 1111, Valpara\'iso, Chile
  \and Department of Physics, University of Oxford, Denys Wilkinson Building, Keble Road, Oxford OX1 3RH, UK
  \and National Centre for Nuclear Research, ul. Pasteura 7, 02-093 Warsaw, Poland
  \and Scuola Normale Superiore, Piazza dei Cavalieri 7, I-56126 Pisa, Italy
  \and The Caltech Optical Observatories, California Institute of Technology, Pasadena, CA, 91125, USA
  }

   \date{Accepted November 30, 2022}
    
  \abstract
   {GDS J033218.92-275302.7 (here GS-14) is a $z\sim5.5$ galaxy with unusual UV spectral features that have been interpreted as signatures of either a double stellar population or of an active galactic nucleus (AGN). GS-14 was detected in \Cii\ as part of the ALPINE survey, turning out to be the galaxy with the lowest molecular gas fraction ($f_{mol}=M_{\rm{molgas}}/(M_{\rm{molgas}}+M_*)\sim0.1$) of that sample.}
   {We exploited the multi-wavelength coverage of GS-14 to investigate the properties and the origin of its emission.}
   {We performed UV-to-NIR SED-fitting, with single/double stellar population and/or AGN component. We analyzed the latest release of the VIMOS spectrum, which shows highly-ionized emission lines (\Ovi, \Nv, and \Niv). The line equivalent widths and line ratios have been compared with those observed in galaxies and AGN, as well as with the predictions from radiation transfer models for star-forming galaxies, AGN, and shocks.}
   {The SED-fitting provides a total stellar mass of $M_*=(4 \pm 1) \times 10^{10}\,\rm{M_\odot}$, an age of the main stellar population of $\sim670\,\rm{Myr}$ and a recent short ($\sim 8\,\rm{Myr} $) burst of star formation (SF) of $\sim 90\,\rm{M_\odot\,yr^{-1}}$. We do not find a significant contribution from an AGN, although we do not have any coverage in the mid-IR, where the dust emission of the AGN would peak. The \nv\ line has a characteristic P-Cygni profile. Fitting it with stellar models suggests a $\sim 3\,\rm{Myr}$ old population of stars with a mass of $\sim 5 \times10^{7}\,\rm{M_\odot}$, consistent with a second component of young stars, as found in the SED-fitting analysis. The \nv\ profile also shows evidence for an additional component of nebular emission. The comparison of the line ratios (\niv/\nv\ and \ovi/\nv) with theoretical models allows us to associate the emission with SF or AGN, but the strong radiation field required to ionize the \ovi\ is more commonly related to AGN activity.}
   {Studying GS-14, we found evidence for an old and already evolved stellar population at $z\sim 5.5$ and show that the galaxy is experiencing a second short burst of SF. In addition, GS-14 carries signatures of obscured AGN activity. The AGN could be responsible for the short depletion time of this galaxy, thus making GS-14 one of the two ALPINE sources with hints of an active nucleus and an interesting target for future follow-ups to understand the connection between SF and AGN activity.}  

\maketitle
\section{Introduction}
It is established that the growth of supermassive black holes (SMBH) and the evolution and properties of their host-galaxies must be coupled, but the exact mechanism, timescales, and how they influence each other are still debated. The presence of scaling relations between the mass of the SMBH and several physical properties of the host galaxy, like galaxy bulge mass, luminosity, and velocity dispersion (e.g. \citealt{kormendy95,magorrian98,ferrarese00,gebhardt00,kormendy13}) led to the formulation of the AGN-galaxy co-evolution paradigm (e.g. \citealt{hopkins08}). In this co-evolution scenario, on the one hand, the stellar feedback may help in funneling gas towards the  nuclear region of the galaxy, thus triggering the Active Galactic Nucleus (AGN) activity. On the other hand, the AGN feedback heats up and expels the gas \citep{zana22}, reducing or quenching the star-formation (SF). Signatures of this interaction have been observed in winds and outflows of cold molecular gas (e.g. \citealt{feruglio10,cicone14}), neutral atomic \citep{rupke05}, and ionised \citep{weymann91,mckernan07} gas. The interplay between AGN and SF of main-sequence (MS) galaxies has been studied in the local universe and at the \textit{``cosmic noon''} ($z\sim2$), but scarcely at high redshift ($z\sim 5-6$), where most of our knowledge comes from luminous quasars (QSO) and starburst (SB) galaxies (e.g. \citealt{bischetti22}).   
A key step of the co-evolution scenario is represented by the obscured-accretion phase, where most of the stellar mass formation and of the BH accretion should take place. The main difficulty in studying this phase is that the AGN activity is shrouded by gas and dust and likely embedded in the galaxy emission, thus, difficult to identify. This is particularly true for low-luminosity AGN (LLAGN): which have their mid-IR emission diluted and overshadowed by the host-galaxy luminosity \citep[i.e. ][]{gruppioni13}. In addition, even the  most recent X-ray facilities struggle to detect high-$z$ LLAGN, due to their low X-ray photon fluxes \citep{barchiesi21}.\par 
High-redshift galaxies with ongoing SF and indication of a possible AGN are the sources to focus on when searching for hidden BH growth as well as for the impact of AGN on the SF (hence early galaxy evolution) itself. GDS J033218.92-275302.7 (hereafter GS-14) falls exactly within those parameters.
\par
\begin{figure}
  \centering
  \resizebox{0.75\hsize}{!}{\includegraphics{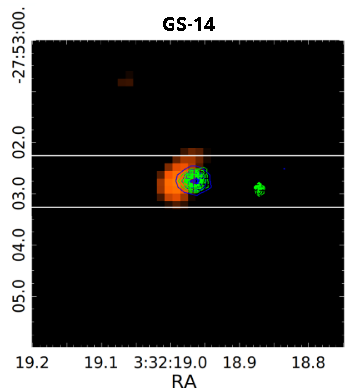}}
  \caption{GS-14 \Cii\ flux map. The \textit{green contours} represent the emission in the \textit{r+} SUBARU filter where the \Lyalpha\ line falls in, the \textit{blue contours} the emission in the ULTRAVISTA Ks filter\citep{cassata20}; the source is not resolve in all the three wavelength ranges. \citet{vanzella10} associated an effective radius to the Ks emission of $r_e[3300\AA]<0.9\,$kpc.}
  \label{fig:morpho} 
\end{figure}

GS-14 is a $z=5.56$ \citep{raiter10,vanzella10} MS-galaxy with unusual spectral features, which have been interpreted as signatures of a double stellar population or linked to AGN activity. The nature of GS-14 emission has been extensively debated since its discovery in the southern field of the Great Observatories Origins Deep Survey (GOODS) during the ESO/FORS2 survey \citep{vanzella06}. \citet{fontanot07} selected this source as a QSO candidate on the basis of its $z_{850}$ magnitude and color selection, but reclassified it as an H II star-forming galaxy due to the presence of the \Niv\ and the lack of the \Nv\ lines in the FORS2 spectrum \citep{vanzella06}. GS-14 appears as a compact-source at all wavelengths (Fig.~\ref{fig:morpho}), altought marginally resolved both in the $i_{775}$ and $z_{850}$ bands \citep{vanzella10}.  In \citet{wiklind08} the source is classified as a Balmer galaxy, with the discontinuity between the \textit{Ks} and $3.6\,\mu$m filter indicating an evolved stellar population.
In \citet{vanzella10}, the bright \Lyalpha\ line and the detection of \niv\ emission line have been interpreted as signature of a young population of massive stars or, alternatively, of an AGN. They also performed Spectral Energy Distribution (SED)-fitting using 16 photometric bands from UV to near-IR (NIR), and found that the source may host a double stellar population, composed of an evolved/aged population, and of a young one of massive stars. \citet{grazian20} deeply analyzed the same FORS2 spectrum of \citet{vanzella10}, and new VIMOS and X-SHOOTER spectra of this source. In addition to the two lines detected by \citep{vanzella10}, they also found \Ovi, and \Nv\ emission lines, which led them to regard GS-14 as an AGN. Regarding the X-ray bands, GS-14 is undetected in the ultradeep 7 Ms X-ray image by Chandra with a flux limit of $10^{-17}\, \rm{erg\,cm^{-2}\, s^{-1}}$ in the observer-frame $0.5-2.0$ keV band \citep{giallongo19}. \par
GS-14 has also been selected as part of the ALMA Large Program to INvestigate C+ at Early Times survey, (ALPINE, \citealt{lefevre20, faisst20, bethermin20}) as a normal star-forming galaxy (SFG). It has been detected in the \Cii\, emission line with a Signal-to-Noise ratio ($S/N$) of $4.6$, but it is not spatially resolved (with a beam of $0.7''$, $\sim4.2\,$kpc, Fig.~\ref{fig:morpho}). There is no detection of continuum near \cii\, with an upper limit of $L_{\rm{IR}} < 2.0 \times 10^{10}\,\rm{L_\odot}$ in the far-infrared dust emission \citep{bethermin20}. With a molecular gas fraction of $\rm{f}_{\rm{mol}}=M_{\rm{mol}}/(M_*+M_{\rm{mol}})=0.10_{-0.06}^{+0.13}$ (derived from the \cii\ luminosity, \citealt{mirka20}), GS-14 is the \cii-detected source with the lowest  $\rm{f}_{\rm{mol}}$ of the entire ALPINE sample.

In this work we exploit the multi-wavelength spectro-photometric coverage of GS-14 to investigate the physical origin of its emission. In §\ref{sec:spec} we report the spectral analysis of GS-14, and compare the observed lines with sources from the literature and with model predictions. The photometric analysis and SED-fitting is reported in §\ref{sec:photom}. We discuss our results and draw the conclusions in §\ref{sec:conclusion}. Throughout this paper, we adopt a \citet{chabrier03} initial stellar mass function and the following cosmological parameters: H$_0 = 70\, \text{km}\, \text{s}^{-1}\, \text{Mpc}^{-1}$, $\Omega_{\text{M}} = 0.3$ and $\Omega_{\Lambda} = 0.7$ \citep{spergel03}.

\section{Spectral Analysis}\label{sec:spec}
\subsection{Spectroscopic data}
Different rest-UV spectroscopic observations are available for GS-14 in terms of depth and resolution: a 4 hours FORS2 spectrum \citep{vanzella10}, a 20 hours VIMOS spectrum (ID 194.A-2003, \citealt{mclure18}), and a 49 hours X-SHOOTER spectrum obtained under two observing programs (384.A-0886 and 089.A-0679). We focus on the VIMOS spectrum, as it has the best S/N among the three available spectra, and it allows for the first time to detect the rest-frame UV continuum of this galaxy at $z \sim 5.5$ at a $S/N=3.7$ (as computed in the $8200-9200\,\AA$ observed-frame wavelength range). \par
\begin{figure*}
  \centering
  \resizebox{\hsize}{!}{\includegraphics{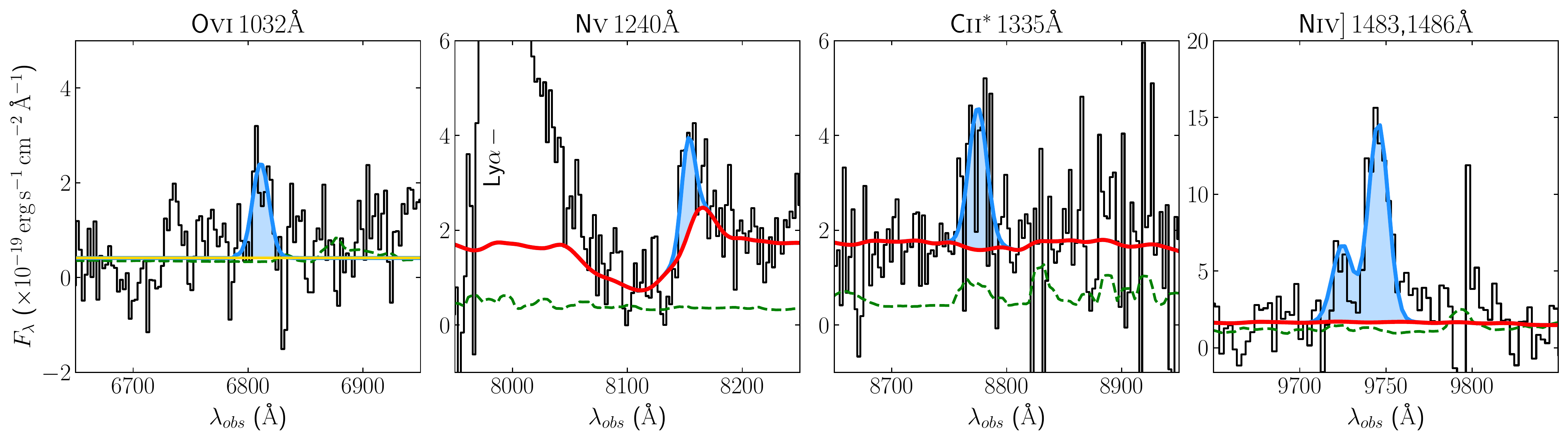}}
  \caption{Spectra and best fit models for the GS-14 main spectral features analysed in this work. From left to right: \Ovi, \Nv, \Ciistar, and \Niv. The \textit{red line} is the SS99 best fit model, the Gaussian best fit for the emission feature is plotted in \textit{blue}, the \textit{yellow line} represents the continuum in the Lyman forest region, and the \textit{green line} the spectrum noise. In the second panel, the strong emission line at $\lambda_{\rm{obs}}\sim8000\,\AA$ is the Ly$\alpha$. For the complete VIMOS spectrum, see Figure 2 in \citet{grazian20}.}
  \label{fig:vimos_spectrum} 
\end{figure*}

Figure~\ref{fig:vimos_spectrum} shows four windows of the one dimensional spectrum of GS-14. The main features of the spectrum are the \Lyalpha\ line and the almost total intergalactic medium (IGM) absorption blueward of it, at wavelength $\lambda_{\rm{obs}} < 8000\,\AA $. \Nv, and \Niv\ are detected at a $S/N$ of $4.6$ and $10.1$, respectively ($\,S/N=F_{\rm{line}}\,/ (\int_{\lambda_0-3\sigma}^{\lambda_0-1.5\sigma} N_{\lambda}\, d\lambda\,+\int_{\lambda_0+1.5\sigma}^{\lambda_0+3\sigma} N_{\lambda}\, d\lambda\,)\, $, where $F_{\rm{line}}$ is the line integrated flux, $N_{\lambda}$ is the spectrum noise, $\lambda_0$ and $\sigma$ are the line centroid and width). \Ovi\ and \Ciistar\ are detected at a $S/N$ of $4.4$ and $4.0$, respectively. \Civ\ is not detected, as this line falls at the edge of the VIMOS spectral range and over a sky emission line. 

\subsection{Spectral fitting}\label{sec:spec_fit}
The latest release of the VIMOS spectrum reaches a continuum $S/N$ of $3.7$, high enough to show for the first time a clear P-Cygni \nv\ line profile. P-Cygni profiles in \nv\ are not unheard of \citep{jaskot17, vanzella18, matthee22} and can be produced by stellar winds of very young and massive stars (e.g. \citealt{prinja86}) or mimicked by Broad or Narrow Absorption Line quasar (BAL/NAL QSO, \citealt{bentz14, appenzeller15}). As our spectrum does not show any absorption line and there are no signs of GS-14 being a type 1 QSO (e.g. $>1000\,\rm{km\,s^{-1}}$ broad emission lines or high X-ray luminosity), the origins of the P-Cygni profile is likely associated with a young population of massive stars. We fitted the spectrum with the models from the STARBURST99 (S99) spectral synthesis code \citep{leitherer99}, adopting the same methodology as \citet{rui21}. S99 models allow the fitting of the stellar continuum and of the lines produced in the stellar atmosphere but not lines that are produced in the ionized gas in the interstellar medium (ISM) of the galaxy. High-resolution UV model spectra with burst ages in the $0.01-20\,$Myr range and metallicities between $0.03$ and $0.5\,\rm{Z_{\odot}}$ were rebinned and smoothed to the VIMOS spectrum resolution. To take into account dust attenuation, the \citet{calzetti00} extinction curve was used to match the \ebv\ of the models with the one measured on the spectrum of GS-14 (\ebv$=0.05$), from the $9200-12000\AA$ observed-frame wavelength range). The free-from-lines spectral window $8400-8600\AA$ was used to normalize the flux of the S99 models to the GS-14 flux. Finally, we performed a $\chi^2$ minimization on the $8050-8400\AA$ range to find the best fit for the \nv\ line profile. We found that the VIMOS spectrum of GS-14 is best fitted with a $2.7 \pm 0.1\,\rm{Myr}$ old stellar population, a mass of $(5\pm 1)\times10^{7}\,\rm{M_{\odot}}$, and a metallicity of $0.5\,\rm{Z_{\odot}}$. We note that the fit of the P-cygni exhibits a bit of degeneracy with the metallicity: the \nv\ P-cygni profile varies little with different values of metallicity, while it is extremely sensitive to the age of the stellar population \citep[see][]{rui21}. Our best-fit model is able to reproduce the absorption part of the \nv\ line profile, as well as the continuum redwards of the \Lyalpha\ emission line. However, the best-fit S99 model is not able to fully reproduce the observed emission in \nv\ and we need an additional emission component. The Gaussian fit of this component provides an emission line centered at $8152.6_{-0.1}^{+0.6}\,\AA$, with a full width half maximum of \textit{FWHM}$=13.9_{-1.9}^{+2.4}\,\AA$ and a flux of $2.9_{-0.1}^{+0.9}\,\times10^{-18}\,\rm{erg\,cm^{-2}\,s^{-1}}$.\par
We also fitted the \ovi, \ciistar, and \niv\ emission lines using Gaussian components. Their best-fit values are reported in Table~\ref{tab:lines}. Due to the IGM absorption, the measured \ovi\ flux should be considered as a lower-limit. \par
We compared the EW of the \nv\ line in absorption and emission with the Binary Population and Spectral Synthesis code (BPASS, \citealt{stanway18}) and with the S99 models. We find that there are models able to reproduce the observed EW in absorption and that those have a very young stellar population, in agreement with what we find from the P-Cygni \nv\ line profile. However, regarding the emission component, all the models have EW$>-3\,\AA$, far from the measured $EW_{\rm{\nv},em}=-5.2\,\AA$. This result tells us that there are likely two different mechanisms at the origin of the \nv\ emission: stellar winds from young and massive stars create the absorption component and contribute to the emission,  while an additional emission, not linked to stellar winds phenomena, provides the rest of the observed \nv\ flux. This additional contribution is likely due to the ionized gas in the ISM, which are not modelled by stellar synthesis codes such as BPASS or S99.

\begin{table}
\caption{Properties of the emission lines of GS-14, derived from the VIMOS spectrum, fitting the lines with a Gaussian profile. $\lambda_{0}$ refers to the observed-frame centroid, \textit{FWHM} to the Full Width Half Maximum, and $F$ to the line flux. For the \nv\ emission line, the values refers to the fit obtained once the stellar component of the P-Cygni profile has been subtracted. The \ovi\ line lies in the Lyman forest and its continuum is not detected, therefore its flux should be considered a lower limit.}
\label{tab:lines}
\centering
\begin{tabular}{cccc}
\hline \hline
Line & $\lambda_{0}$ & \textit{FWHM} & $F$ \\[3pt]
     &  $\AA$   & $\AA$ & $\rm{erg\,s^{-1}\,cm^{-2}}$ \\
\hline
\Ovi & $6811.0_{-1.0}^{+0.1}$ & $17.3_{-2.2}^{+4.1}$ & $>3.6_{-1.2}^{+0.9} \times 10^{-18}$ \\[3pt]
\Nv  & $8152.6_{-0.1}^{+0.6}$ & $13.9_{-1.9}^{+2.4}$ & $2.9_{-0.1}^{+0.9} \times 10^{-18}$ \\[3pt]
\Ciistar &  $8775.0_{-1.0}^{+1.0}$ & $19_{-5}^{+14}$ & $6.0_{-3.0}^{+4.5} \times 10^{-18}$ \\[3pt]
N\,\textsc{iv}]\,$1483\AA$ & $9724.8_{-0.1}^{+0.4}$ & $13.9_{-1.9}^{+2.4}$ & $7.3\pm1.3\,\times 10^{-18}$ \\[3pt]
N\,\textsc{iv}]\,$1486\AA$ & $9745.6_{-0.1}^{+0.4}$ & $13.9_{-1.9}^{+2.4}$ & $1.9\pm 0.3 \,\times 10^{-17}$ \\[3pt]
\hline
\end{tabular}
\end{table}

\subsection{Comparison with literature}\label{sec:literature_comparison}
We performed a deep search for sources in the literature with \Ovi, \Nv, or \Niv\ emission lines; our goal is to discern the nature (SF or AGN) of GS-14. We found eight sources with all the three lines, seven from \citet{dietrich03} at redshift $3.9<z<5$ and one from \citet{baldwin03} at $z=1.96$, all of these are spectroscopically confirmed AGN. The three lines can also be found in the  \citet{hainline11} composite spectrum of 33 narrow-line AGN at $z\sim 2-3$. In the collection of sources, we also checked if the flux of the \niv\ was higher than that of the \nv, as it is the case for GS-14. None of the above-mentioned AGN shows this characteristic. We note that the $z=3.36$ lensed galaxy of \citet{fosbury03} and the stacked spectrum of $z\sim2-3.8$, EW$_{\rm{C\textsc{iii}]}}\geq20\AA$ galaxies of \citet{lefevre19} show a \niv\ emission line more luminous than the \nv, but both lack \ovi\ detection. With the exception of the young SFG of \citet{rui21} (which shows P-Cygni profiles in both \ovi\ and \nv, but no \niv\ emission line detection), all the sources with \nv\ and \ovi\ detections are AGN. Being an high-ionization emission line, the \ovi\ line is usually associated with AGN activity, although it has also been detected in emission in a small number of extreme SFG \citep{otte03, grimes07, matthew16}.  We report in Appendix~\ref{sec:app_sources} the complete list of the sources we compared with, as well as the detected lines.   

\subsection{Comparison with theoretical predictions}
\begin{figure}
  \centering
  \resizebox{\hsize}{!}{\includegraphics{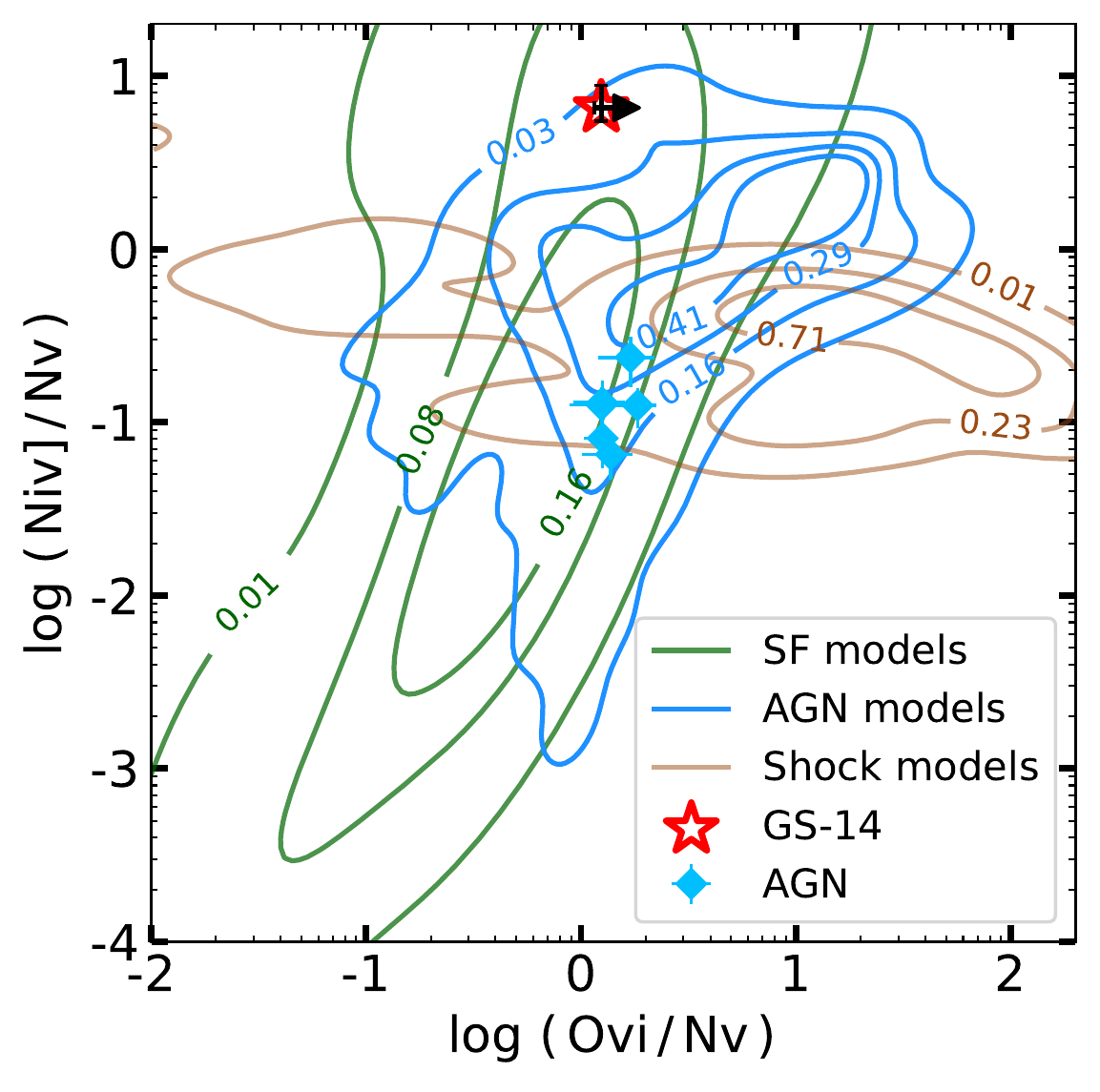}}
  \caption{Comparison of GS-14 (\textit{red star}), literature AGN (\textit{cyan diamonds}), and nebular theoretical predictions of the \niv/\nv\ and \ovi/\nv\ flux ratios. The contours represents iso-proportions of the density (i.e. 8\% of the probability mass lies outside of the contour drawn for 0.08). \textit{Brown contours} refer to shock predictions from the \citet{morisset15} 3MDB$\rm{^S}$ database, \textit{green contours} to SF models from \citet{gutkin16}, and \textit{blue contours} to AGN emission as computed by \citet{feltre16}. Literature AGN are the sources described in §\ref{sec:literature_comparison} and Appendix~\ref{sec:app_sources} with all the three lines detected. No shock model is able to reproduce the GS-14 line ratios; both SF and AGN can be at the origin of the observed nebular emission. As the GS-14 \ovi\ line is not corrected for the IGM attenuation, we have only a lower limit on its flux, thus a lower limit on the \ovi/\nv\ ratio (\textit{black arrow}).}
  \label{fig:line_ratio} 
\end{figure}

We exploit the \niv/\nv\ and \ovi/\nv\ flux ratios to further investigate the origin of the additional \nv\ emission component, as well as  \niv\ and \ovi\ emission lines. Figure~\ref{fig:line_ratio} shows the comparison of the line ratios in GS-14 (\textit{red star}) with those from the AGN of §\ref{sec:literature_comparison} with all the three lines detected (\textit{cyan diamonds}). GS-14 \niv/\nv\ line ratio uncertainty is shown as a \textit{black error bar}, and the \ovi/\nv\ lower limit with the \textit{black arrow}. The contours refer to the theoretical predictions for flux ratios driven by shocks, AGN, and SF. The shock predictions (\textit{brown contours}) are from the Mexican Million Models Database (3MDB$\rm{^S}$, \citealt{morisset15}), a compilation of shock models calculated with the code MAPPINGS and evaluated with the CLOUDY \citep{ferland13} photoionization code. The SF models (\textit{green contours}) are from \citet{gutkin16} models, and refer to a cloud illuminated by a $10\,\rm{Myr}$-old population, with the nebular emission computed with CLOUDY. Finally, the AGN predictions (\textit{blue contours}) are from \citet{feltre16} and rely on CLOUDY to simulate the emission from a Narrow Line Region (NLR) cloud illuminated by the central AGN. We find that there is no shock model able to reproduce the observed flux ratios of GS-14, while both SF and AGN models are compatible with the observed values. We note that, as we must consider the \ovi\ flux as a lower limit, in GS-14 the intrinsic \ovi/\nv\ flux ratio should be higher; in fact, the IGM attenuation at \ovi\ wavelength for a $z\sim5.5$ source could be as high as a factor 4 (e.g. \citealt{inoue14}) and GS-14 should move towards the region where less SF models reside. Table \ref{tab:param_space_lineratio} reports the parameter space explored by the shock, AGN, and SF models we used. While we cannot constrain the ionization parameter using the GS-14 flux ratios, we find that, for the AGN models, only hydrogen density of the cloud of $n_H\sim10^4\,\rm{cm}^{-3}$ and metallicity $Z\leq0.5\,\rm{Z_{\odot}}$ fit our data. Regarding the SF models, the observed line ratio can be reproduced only by $Z\leq0.13\,\rm{Z_{\odot}}$ models. With such low-metallicity, stellar models predict much weaker P-Cygni profile in the \nv\ wind line than the observed one. This suggests that the nebular emission in GS-14 is likely powered by an AGN. In the future, observing more emission lines, like \Civ\ and \Heii, could provide a definitive answer on the origin of the emission.

\begin{table}
\caption{Parameter space for the AGN, SF, and shock models. For each parameter, \textit{N$_{\rm{sample}}$} values are simulated in the \textit{values} range. \textit{U$_S$} refers to the ionization parameter at the Str\"omgren radius (see \citet{feltre16}), $Z$ to the metallicity, $\xi_{\rm{d}}$ to the dust-to-metal mass ratio, $n_{\rm{H}}$ to the cloud gas density, and $\alpha$ to the power law index at UV and optical wavelengths of the AGN continuum ($S_{\nu} \propto \nu^{\alpha}$ for $0.001 \leq \lambda/\mu\rm{m} \leq 0.25$). For the shock models, $v_{\rm{shock}}$ refers to the shock velocity, $n_{\rm{H,pre-shock}}$ to the pre-shock density, and $B_{\rm{transv}}$ to the transverse magnetic field; the investigated metallicity parameter space is the same of the SF models.}
\label{tab:param_space_lineratio}
\centering
\begin{tabular}{cccc}
& & N$_{\rm{sample}}$ & values \\
\hline \hline
AGN & $\log \rm{U}_{S}$ & 5 & $[-5,-1]$  \\
& $Z/\rm{Z_{\odot}}$ & 16 & $[0.06,4.6]$ \\
& $\xi_{\rm{d}}$ & 3 & $0.1,0.3,0.5$\\
& $\log\, (\,\rm{n}_{\rm{H}}/\rm{cm^{-3}}\,)$ & 3 & $2, 3, 4$\\
& $\alpha$ & 4 & $[-2,-1.2]$ \\
\hline
SF & $\log \rm{U}_{S}$ & 7 & $[-1.0,-4.0]$ \\
& $Z/\rm{Z_{\odot}}$ & 14 & $[0.006,2.6]$ \\
& $\xi_{\rm{d}}$ & 1& 0.3\\
& $\log\, (\,\rm{n}_{\rm{H}}/\rm{cm^{-3}}\,)$ & 1 & 2\\
\hline
Shock & $v_{\rm{shock}}/\rm{km\,s^{-1}}$ & 37 & $[100,1000]$ \\
& $\log\,(\,n_{\rm{H,pre-shock}}/\rm{cm^{-3}})$ & 7 & $[0,4]$ \\
& $B_{\rm{transv}}\,/\,\mu\rm{G}$ & 8 & $[10^{-4},10]$\\
& $Z/\rm{Z_{\odot}}$ & & $[0.006,2.6]$ \\
\hline

\hline
\end{tabular}
\end{table}

\section{Photometric analysis}\label{sec:photom}

\subsection{Photometric data}
GS-14 has been observed in several photometric bands. It has been detected with the \textit{MPG-ESO WFI} \citep{hildebrandt06,erben05}, with \textit{ISAAC} instrument at \textit{VLT} \citep{nonino09,wuyts08,retzlaff10}, with \textit{CFHT WIRCam} \citep{hsieh12}, and with the \textit{SUBARU Suprime-Cam} \citep{cardamone10}. Regarding observations from space telescopes, GS-14 has detections with the \textit{ACS} and \textit{WFC3} on board of \textit{HST} \citep{giavalisco04,koekemoer11,grogin11A,brammer12,vandokkum13}, as well as with the \textit{Spitzer IRAC} \citep{dickinson03,ashby13,guo13}. The complete list of photometric filters adopted is reported in Appendix~\ref{sec:app_filters}. We refer to \citet{faisst20} for the in-depth description of the photometric data.\par
The SED of GS-14 is characterized by the evident discontinuity between the \textit{VLT/ISAAC} and \textit{CFHT/WIRCAM} bands and the \textit{Spitzer/IRAC} bands that is consistent with being originated from the Balmer break (see Fig.\ref{fig:sed}). Another peculiarity is the higher flux in the \textit{IRAC 1} and \textit{2} bands compared to the fluxes of bands \textit{3} and \textit{4}. This should be due to a significant contribution of the H$\beta$+\oiii, and H$\alpha$ lines to the \textit{IRAC 1} and \textit{2} fluxes.

\subsection{SED-fitting}\label{sec:sedfitting}
To estimate the galaxy properties, we choose to use the \texttt{X-CIGALE} code \citep{yang20,yang22}, the latest version of the Code Investigating GALaxy Emission (\texttt{CIGALE}, \citealt{burgarella05,noll09,boquien19}) SED-fitting code, as it is quite flexible and allows us to fit the photometric data with and without the AGN component, as well as with a double stellar population. We use the \citet{bruzual03} population synthesis model with a \citet{chabrier03} initial mass function, nebular emission lines, a \citet{calzetti00} dust attenuation law  and the \citet{draine14} dust models. We test different star-formation histories (SFHs): double exponential, delayed SFH, delayed SFH plus burst or quench, and constant. The AGN emission is added via the SKIRTOR models \citep{stalevski12,stalevski16}, in which the torus is modelled as a clumpy two-phases medium \citep[we refer to][for further details]{yang20}. We report in Table \ref{tab:param_space_sed} the complete parameter space investigated via the SED-fitting.   
\par
\begin{figure}
  \centering
  \resizebox{\hsize}{!}{\includegraphics{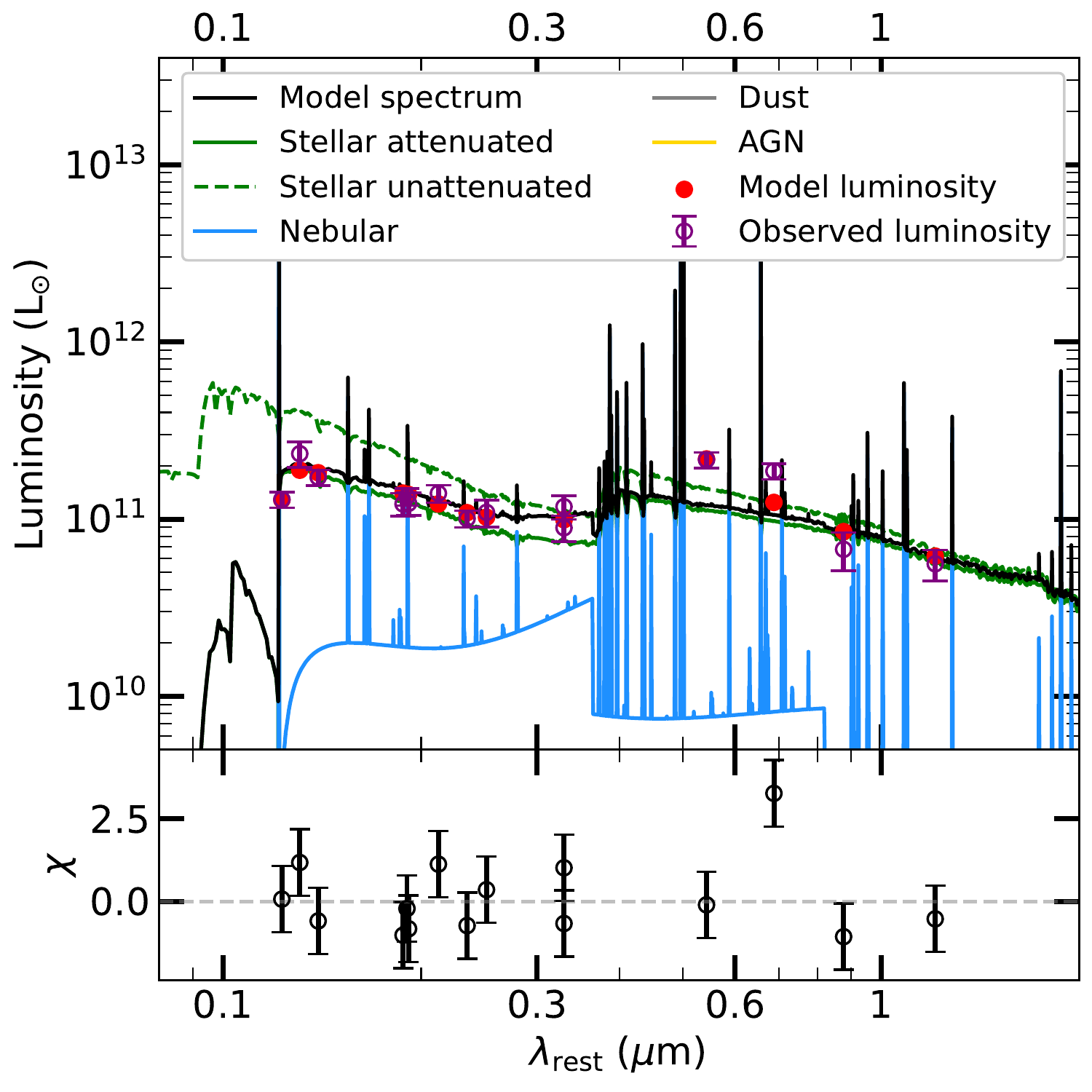}}
  \caption{\texttt{X-CIGALE} best-fit SED of GS-14. The fitting prefers a  delayed exponential (with optional exponential burst) SFH, with an old ($t_{\rm{age}}=680\pm 170\,\rm{Myr}$) stellar population of $M_*=(4 \pm 1) \times 10^{10}\,\rm{M_\odot}$, and a young one of $8\pm6\,\rm{Myr}$, which is now experiencing a burst of SF of $ 90\pm 30\,\rm{M_\odot\,yr^{-1}}$. The dust and AGN component are not visible in this plot, as they contribute the most at $\lambda_{\rm{rest}}>3\,\mu$m. The investigated parameter space is reported in Table~\ref{tab:param_space_sed}.}
  \label{fig:sed} 
\end{figure}

Figure~\ref{fig:sed} shows the best-fit model for GS-14, with a delayed SFH, a stellar mass of $M_*=(4 \pm 1) \times 10^{10}\,\rm{M_\odot}$, a stellar age of $680\pm 170\,\rm{Myr}$,  and a burst of SF of $\rm{SFR}=90\pm 30\,\rm{M_\odot\,yr^{-1}}$ in the last $8\pm6\,\rm{Myr}$. We find that the stellar mass, age of the galaxy and bulk of the SFH are well constrained and do not depend heavily on the presence of AGN or of a double population. Although the fitting is acceptable with a single population ($\tilde{\chi}^2\sim 2.7$), we obtain a better fit when adding a recent burst in the SFH ($\tilde{\chi}^2= 1.6$). In particular, the presence of a younger population allows us to better fit the higher fluxes of the \textit{IRAC 1} and \textit{2} bands (with respect to \textit{IRAC 3} and \textit{4}) due to the H$\beta$ and H$\alpha$ lines of the young stars contributing significantly to the \textit{IRAC 1} and \textit{2} fluxes. \par
The fitting does not reveal any significant contribution from an AGN. When AGN models are used, the SED-fitting constantly prefers low-luminosity type 2 AGN models, with an AGN contribution to the optical-UV six order of magnitude lower than the stellar emission. We note, however, that we do not have any coverage in the mid-IR, where the warm dust heated  by the AGN should contribute the most. The SED-fitting, nevertheless, allows us to exclude the presence of a type 1 AGN, as it would contribute significantly in the optical-UV where we have an optimal photometric coverage, but leaves open the possibility of a moderate- or low-luminosity obscured AGN. This is consistent with the narrow line profiles observed in GS-14 (\Lyalpha, \nv).\par
As sanity check, we also perform SED-fitting with the SED3FIT code \citep{berta13,cuna08}, although without the option of a double population. We find that the stellar mass and SFR are compatible with those from \texttt{X-CIGALE} within the uncertainties. And similarly, no significant AGN contribution is detected. 

\begin{table}
\caption{GS-14 properties from photometric and spectral analysis. $M^{\rm{tot}}_*$, $M^{\rm{old}}_*$, and $M^{\rm{young}}_*$ refers to the total, the old, and the young population, respectively. \textit{age}$^{\rm{old}}_*$ and \textit{age}$^{\rm{young}}_*$ to the age of the old and young stellar population. \textit{SFR} refers to the instantaneous SFR from the SED-fitting, \textit{SFR}$^{\rm{MS}}$ to the expected SFR for a MS galaxy of similar mass and redshift, obtained using the \citet{speagle14} MS, \textit{SFR}$^{\rm{\cii}}$ refers to the SFR estimated from the \cii\ luminosity and using the \citet{schaerer20} relations of Figure 3, and \textit{SFR}$^{\rm{UV}}$ to the SFR obtained from the UV luminosity. $Z$ refers to the metallicity. The molecular gas fraction $f_{\rm{mol}}$ was computed with the $M_{\rm{molgas}}$ of \citet{mirka20}. The \ebv\ in the \textit{Photometric column} was derived from the SED-fitting by \citet{faisst20}, while the one in the \textit{Spectral column} from the VIMOS spectrum as described in §\ref{sec:spec_fit}. $L_{\rm{bol,AGN}}^{\rm{SED}}$ and $L_{\rm{bol,AGN}}^{\rm{X-ray}}$ are the AGN bolometric luminosity derived from the SED-fitting and from the \textit{Chandra} X-ray flux upper limit, assuming the bolometric correction from \citet{lusso12}, respectively.}
\label{tab:recap}
\centering
\begin{tabular}{cccc}
\hline\hline
& Photometric & Spectral \\[3pt]
\hline
$\log (M_*^{tot}/\rm{M_\odot})$ & $10.62_{-0.13}^{+0.10}$ & \\[3pt]
$\log (M_*^{old}/\rm{M_\odot})$ & $10.61_{-0.14}^{+0.10}$ &\\[3pt]
$\log (M_*^{young}/\rm{M_\odot})$ & $8.7\pm0.1$ & $7.7_{-0.2}^{+0.1}$ \\[3pt]
\textit{age}$_*^{\rm{old}}\,$(Myr) & $680 \pm 170$ & \\[3pt]
\textit{age}$_*^{\rm{young}}\,$(Myr) & $8\pm6 $ & $2.7 \pm 0.1$ \\[3pt]
\textit{SFR} (M$_\odot/\rm{yr})$ & $90 \pm 30$ &  \\[3pt]
\textit{SFR}$^{\rm{MS}}$ (M$_\odot/\rm{yr})$ & $\sim200 $ &  \\[3pt]
\textit{SFR}$^{\rm{\cii}}$ (M$_\odot/\rm{yr})$ & &$ 16_{-7}^{+14} $ \\[3pt]
\textit{SFR}$^{\rm{UV}}$ (M$_\odot/\rm{yr})$ & $26 \pm5 $ &  \\[3pt]
$Z/\rm{Z_{\odot}}$ & $<0.2$ & 0.5\\[3pt]
$f_{\rm{mol}}$ & $0.10_{-0.06}^{+0.13}$ & \\[3pt]
\ebv & 0.05 & 0.05\\[3pt]
$\log (L_{\rm{bol,AGN}}^{\rm{SED}}/\rm{erg\,s^{-1}})$ & <44.5 & \\[3pt]
$\log (L_{\rm{bol,AGN}}^{\rm{X-ray}}/\rm{erg\,s^{-1}})$ & <43.5 & \\[3pt] 
\hline
\end{tabular}
\end{table}

\section{Discussion and conclusions}\label{sec:conclusion}
We find several clues indicating that GS-14 has a double stellar population. The P-Cygni profile in the \nv\ line suggests the presence of a young population of massive stars. The fitting of the line profile provides an age of $2.7\pm0.1\,\rm{Myr}$ and a mass of $(5\pm1)\times 10^7\,\rm{M_\odot}$. Moreover, the best SED-fitting is obtained for a double population, with a $680\pm 170\,\rm{Myr}$ old population of $(4 \pm 1) \times 10^{10}\,\rm{M_\odot}$ and a young ($8\pm6\,\rm{Myr}$ old) one of $ (5.6\pm0.1) \times 10^8\,\rm{M_\odot} $ (see Table~\ref{tab:recap}). In this scenario, the old population dominates and accounts for most of the stellar mass of the galaxy (similar to what was found by \citealt{laporte21, harikane22, matthee22}). It is responsible for the Balmer break and for the continuum at $\lambda_{\rm{rest}}\gtrsim 3000\,\AA$. The young population is linked to the ongoing SF and produces the P-Cygni profile and most of the continuum at $\sim2000\,\AA$. The low cold-gas fraction of GS-14 of $f_{\rm{mol}}\sim 0.10_{-0.06}^{+0.13}$, derived from the \Cii\ emission (see \citealt{mirka20}) and the SED-fitting stellar mass, favors the scenario of GS-14 being composed mostly of an old and evolved stellar population formed $\sim600\,$Myr after the Big Bang, that has already consumed or expelled most of its original gas reservoir. We exclude the alternative origin of the \nv\ P-Cygni profile, i.e. BAL-QSO absorption feature mimicking a P-Cygni profile, as there is no observational evidence that GS-14 could be a type 1 QSO and the spectra show no signs of absorption features \citep[e.g. \Lyalpha, \Civ, \Ovi, \Siiv;][]{vito22,vietri22}.\par
The specific SFR ($\rm{sSFR}=\rm{SFR}/\rm{M_*}$) of GS-14 is $\log{(\rm{sSFR/yr^{-1}})}=8.7_{-0.2}^{+0.1}$. At this redshift, and considering the mass from SED-fitting, the SFR for a MS galaxy is $\rm{SFR}^{MS}\sim 200\, \rm{M_\odot\,yr^{-1}}$ \citep{speagle14}, with  the lower boundary of its $1\,\sigma$ dispersion at $\sim 60\,\rm{M_\odot\,yr^{-1}}$; hence GS-14 with its $\rm{SFR}=90 \pm 30\,\rm{M_\odot\,yr^{-1}}$, while below the average MS, is still within its dispersion. From the SED-fitting derived SFH, without considering the recent episode of SF, the SFR should be $\sim 1 \, \rm{M}_{\odot}\,\rm{yr^{-1}}$, thus GS-14 would be $>4\,\sigma$ below the MS, in the locus of quiescent galaxies. The new episode of SF has moved GS-14 up towards the MS, altought it should last only a few tens of Myr due to its short depletion time (t$_{\rm{depl}}=\rm{M_{\rm{molgas}}}/\rm{SFR}=80_{-60}^{+130}\,\rm{Myr}$). This kind of up-and-down movement in the SFR-M$_*$ plane is expected: in fact, \citet{tacchella16}, and \citet{orr19} simulations suggest that the MS dispersion can be originated by similar oscillations in sSFR on time-scales $\sim0.4\,t_{\rm{Hubble}}$.
\par
The GS-14 \nv\ line profile shows, on top of the fitted P-Cygni profile, an additional emission component. The comparison of the observed \nv\ EW with \nv\ equivalent widths from BPASS and S99 models, allows us to exclude that the origin of the whole \nv\ emission is from stellar winds. This additional emission component, thus, comes from nebular emission, but it can have various origins, depending on which process originates the radiation field that illuminates the gas. On the one hand, the shock origin can be excluded, as the GS-14 flux ratios of \niv/\nv\ and \ovi/\nv\ are not compatible with shocks-related models (Figure~\ref{fig:line_ratio}). On the other hand, models of AGN or SF are both able to reproduce the observed flux ratios, as shown in Figure~\ref{fig:line_ratio}. However, we note that an AGN origin is more likely, as the high-ionization potential of the \ovi\ requires extreme conditions to be of stellar origin. Moreover, we have found in the literature just one star-forming galaxy \citep{rui21} with both \ovi\ and \nv, all the remaining sources with these two lines are AGN. Finally, all the sources with all the three \ovi, \nv, and \niv\ lines are AGN (Table~\ref{tab:literature_sources}). While none of these is a sufficient proof, they strongly point towards an AGN origin for the GS-14 nebular emission. The lack of X-ray detection for GS-14 in deep \textit{Chandra} data and the fact that the SED-fitting does not show a significant contribution from the AGN reveal that this AGN should be a type 2. Obscured (type 2) AGN are more difficult to detect in the X-rays, but can be identified by their high-ionization lines and/or by their mid-IR emission. Exploiting the X-ray flux upper limit and assuming a power-law spectrum (without AGN obscuration) for the X-ray emission with a photon index of $\Gamma=1.8$ ($F(E)\propto E^{1-\Gamma}$) , we derive a $1\sigma$ upper limit on the rest frame $2-10\,$keV luminosity of the AGN: $\log\,(\,\rm{L}_{2-10\,\rm{keV}}\,/\rm{erg\,s^{-1}})<42.5$. Assuming the bolometric correction of \citet{lusso12}, we have an upper limit on the AGN bolometric luminosity of $\log\,(\,L_{\rm{bol,AGN}}^{\rm{X-ray}}\,/\rm{erg\,s^{-1}})<43.5$. The upper limit on the AGN bolometric luminosity derived from the SED-fitting is $\log\,(\,L_{\rm{bol,AGN}}^{\rm{SED}}\,/\rm{erg\,s^{-1}})<44.5$.
AGN activity could also be a possible explanation for GS-14 extreme low content of molecular gas. A strong past phase of nuclear activity, possibly triggered by the first episode of SF, may have expelled or heated up the molecular gas, hence lowered significantly the f$_{\rm{mol}}$ and quenched the SF. The low-power AGN signature we are now witnessing could be the last remnant of this past, more powerful, AGN activity.\par
Our interpretation paints an intriguing picture of GS-14: this $z\sim5.5$ source is an already evolved galaxy possibly formed $600\,$Myr after the Big Bang that is now experiencing a second burst of SF, and is also carrying signatures of obscured AGN activity.\par
Future observations could shed more light on the AGN contribution in GS-14, especially if those observations would target the mid- and far-IR part of the SED ($60\,\mu\rm{m}-3\,\rm{mm}$ observer-frame), where its emission should peak, or at least could be disentangled from that of SF. GS-14 deserves also additional high-resolution observations with ALMA: the \Cii\ detected but not resolved, and the far-IR continuum detection (which is missing for now) would place more stringent constraints to the SF emission and the AGN contribution. Furthermore, GS-14 is the only ALPINE source with hints of obscured AGN activity, which makes its follow-up with deeper and higher-resolution ALMA observations even more important and rewarding. Finally, deep rest-UV spectroscopy, targeting mid- and high-ionization lines, such as \Civ, \Heii, and \Ciii, and exploiting their line ratio diagram, could provide the final evidence of whether the radiation field in GS-14 is dominated by the AGN activity or by a young stellar population. GS-14 was recently observed with the \textit{NIRSpec} aboard the \textit{James Webb Space Telescope} as part of the GTO program n.1216, and hopefully the mystery of the origin of its emission will be definitively unveiled. 


\begin{acknowledgements}
G.C.J. acknowledges funding from ERC Advanced Grant 789056 ``FirstGalaxies’’ under the European Union’s Horizon 2020 research and innovation programme. A.F. acknowledges the support from grant PRIN MIUR 2017-20173ML3WW\_001. E.I. acknowledge funding by ANID FONDECYT Regular 1221846. M.R. acknowledges support from the Narodowe Centrum Nauki (UMO-2020/38/E/ST9/00077).
\end{acknowledgements}

\bibliographystyle{aa.bst} 
\bibliography{GS14.bib} 

\begin{thebibliography}{103}
\expandafter\ifx\csname natexlab\endcsname\relax\def\natexlab#1{#1}\fi

\bibitem[{{Appenzeller} {et~al.}(2005){Appenzeller}, {Stahl}, {Tapken},
  {Mehlert}, \& {Noll}}]{appenzeller15}
{Appenzeller}, I., {Stahl}, O., {Tapken}, C., {Mehlert}, D., \& {Noll}, S.
  2005, \aap, 435, 465

\bibitem[{{Ashby} {et~al.}(2013){Ashby}, {Willner}, {Fazio}, {Huang}, {Arendt},
  {Barmby}, {Barro}, {Bell}, {Bouwens}, {Cattaneo}, {Croton}, {Dav{\'e}},
  {Dunlop}, {Egami}, {Faber}, {Finlator}, {Grogin}, {Guhathakurta},
  {Hernquist}, {Hora}, {Illingworth}, {Kashlinsky}, {Koekemoer}, {Koo},
  {Labb{\'e}}, {Li}, {Lin}, {Moseley}, {Nandra}, {Newman}, {Noeske}, {Ouchi},
  {Peth}, {Rigopoulou}, {Robertson}, {Sarajedini}, {Simard}, {Smith}, {Wang},
  {Wechsler}, {Weiner}, {Wilson}, {Wuyts}, {Yamada}, \& {Yan}}]{ashby13}
{Ashby}, M.~L.~N., {Willner}, S.~P., {Fazio}, G.~G., {et~al.} 2013, \apj, 769,
  80

\bibitem[{{Ba{\~n}ados} {et~al.}(2016){Ba{\~n}ados}, {Venemans}, {Decarli},
  {Farina}, {Mazzucchelli}, {Walter}, {Fan}, {Stern}, {Schlafly}, {Chambers},
  {Rix}, {Jiang}, {McGreer}, {Simcoe}, {Wang}, {Yang}, {Morganson}, {De Rosa},
  {Greiner}, {Balokovi{\'c}}, {Burgett}, {Cooper}, {Draper}, {Flewelling},
  {Hodapp}, {Jun}, {Kaiser}, {Kudritzki}, {Magnier}, {Metcalfe}, {Miller},
  {Schindler}, {Tonry}, {Wainscoat}, {Waters}, \& {Yang}}]{banados16}
{Ba{\~n}ados}, E., {Venemans}, B.~P., {Decarli}, R., {et~al.} 2016, \apjs, 227,
  11

\bibitem[{{Baldwin} {et~al.}(2003){Baldwin}, {Hamann}, {Korista}, {Ferland},
  {Dietrich}, \& {Warner}}]{baldwin03}
{Baldwin}, J.~A., {Hamann}, F., {Korista}, K.~T., {et~al.} 2003, \apj, 583, 649

\bibitem[{{Barchiesi} {et~al.}(2021){Barchiesi}, {Pozzi}, {Vignali}, {Carrera},
  {Vito}, {Calura}, {Bisigello}, {Lanzuisi}, {Gruppioni}, {Lusso},
  {Delvecchio}, {Negrello}, {Cooray}, {Feltre}, {Fern{\'a}ndez-Ontiveros},
  {Gallerani}, {Kaneda}, {Oyabu}, {Pereira-Santaella}, {Piconcelli}, {Ricci},
  {Rodighiero}, {Spinoglio}, \& {Tombesi}}]{barchiesi21}
{Barchiesi}, L., {Pozzi}, F., {Vignali}, C., {et~al.} 2021, \pasa, 38, e033

\bibitem[{{Bentz} {et~al.}(2004){Bentz}, {Osmer}, \& {Weinberg}}]{bentz14}
{Bentz}, M.~C., {Osmer}, P.~S., \& {Weinberg}, D.~H. 2004, in The Interplay
  Among Black Holes, Stars and ISM in Galactic Nuclei, ed.
  T.~{Storchi-Bergmann}, L.~C. {Ho}, \& H.~R. {Schmitt}, Vol. 222, 515--516

\bibitem[{{Berta} {et~al.}(2013){Berta}, {Lutz}, {Santini}, {Wuyts}, {Rosario},
  {Brisbin}, {Cooray}, {Franceschini}, {Gruppioni}, {Hatziminaoglou}, {Hwang},
  {Le Floc'h}, {Magnelli}, {Nordon}, {Oliver}, {Page}, {Popesso}, {Pozzetti},
  {Pozzi}, {Riguccini}, {Rodighiero}, {Roseboom}, {Scott}, {Symeonidis},
  {Valtchanov}, {Viero}, \& {Wang}}]{berta13}
{Berta}, S., {Lutz}, D., {Santini}, P., {et~al.} 2013, \aap, 551, A100

\bibitem[{{B{\'e}thermin} {et~al.}(2020){B{\'e}thermin}, {Fudamoto}, {Ginolfi},
  {Loiacono}, {Khusanova}, {Capak}, {Cassata}, {Faisst}, {Le F{\`e}vre},
  {Schaerer}, {Silverman}, {Yan}, {Amorin}, {Bardelli}, {Boquien}, {Cimatti},
  {Davidzon}, {Dessauges-Zavadsky}, {Fujimoto}, {Gruppioni}, {Hathi}, {Ibar},
  {Jones}, {Koekemoer}, {Lagache}, {Lemaux}, {Moreau}, {Oesch}, {Pozzi},
  {Riechers}, {Talia}, {Toft}, {Vallini}, {Vergani}, {Zamorani}, \&
  {Zucca}}]{bethermin20}
{B{\'e}thermin}, M., {Fudamoto}, Y., {Ginolfi}, M., {et~al.} 2020, \aap, 643,
  A2

\bibitem[{{Bischetti} {et~al.}(2022){Bischetti}, {Feruglio}, {D'Odorico},
  {Arav}, {Ba{\~n}ados}, {Becker}, {Bosman}, {Carniani}, {Cristiani}, {Cupani},
  {Davies}, {Eilers}, {Farina}, {Ferrara}, {Maiolino}, {Mazzucchelli},
  {Mesinger}, {Meyer}, {Onoue}, {Piconcelli}, {Ryan-Weber}, {Schindler},
  {Wang}, {Yang}, {Zhu}, \& {Fiore}}]{bischetti22}
{Bischetti}, M., {Feruglio}, C., {D'Odorico}, V., {et~al.} 2022, \nat, 605, 244

\bibitem[{{Boquien} {et~al.}(2019){Boquien}, {Burgarella}, {Roehlly}, {Buat},
  {Ciesla}, {Corre}, {Inoue}, \& {Salas}}]{boquien19}
{Boquien}, M., {Burgarella}, D., {Roehlly}, Y., {et~al.} 2019, \aap, 622, A103

\bibitem[{{Borguet} {et~al.}(2012){Borguet}, {Edmonds}, {Arav}, {Benn}, \&
  {Chamberlain}}]{borguet12}
{Borguet}, B. C.~J., {Edmonds}, D., {Arav}, N., {Benn}, C., \& {Chamberlain},
  C. 2012, \apj, 758, 69

\bibitem[{{Brammer} {et~al.}(2012){Brammer}, {van Dokkum}, {Franx},
  {Fumagalli}, {Patel}, {Rix}, {Skelton}, {Kriek}, {Nelson}, {Schmidt},
  {Bezanson}, {da Cunha}, {Erb}, {Fan}, {F{\"o}rster Schreiber}, {Illingworth},
  {Labb{\'e}}, {Leja}, {Lundgren}, {Magee}, {Marchesini}, {McCarthy},
  {Momcheva}, {Muzzin}, {Quadri}, {Steidel}, {Tal}, {Wake}, {Whitaker}, \&
  {Williams}}]{brammer12}
{Brammer}, G.~B., {van Dokkum}, P.~G., {Franx}, M., {et~al.} 2012, \apjs, 200,
  13

\bibitem[{{Bruzual} \& {Charlot}(2003)}]{bruzual03}
{Bruzual}, G. \& {Charlot}, S. 2003, \mnras, 344, 1000

\bibitem[{{Burgarella} {et~al.}(2005){Burgarella}, {Buat}, \&
  {Iglesias-P{\'a}ramo}}]{burgarella05}
{Burgarella}, D., {Buat}, V., \& {Iglesias-P{\'a}ramo}, J. 2005, \mnras, 360,
  1413

\bibitem[{{Calzetti} {et~al.}(2000){Calzetti}, {Armus}, {Bohlin}, {Kinney},
  {Koornneef}, \& {Storchi-Bergmann}}]{calzetti00}
{Calzetti}, D., {Armus}, L., {Bohlin}, R.~C., {et~al.} 2000, \apj, 533, 682

\bibitem[{{Cardamone} {et~al.}(2010){Cardamone}, {van Dokkum}, {Urry},
  {Taniguchi}, {Gawiser}, {Brammer}, {Taylor}, {Damen}, {Treister}, {Cobb},
  {Bond}, {Schawinski}, {Lira}, {Murayama}, {Saito}, \&
  {Sumikawa}}]{cardamone10}
{Cardamone}, C.~N., {van Dokkum}, P.~G., {Urry}, C.~M., {et~al.} 2010, \apjs,
  189, 270

\bibitem[{{Cassata} {et~al.}(2020){Cassata}, {Morselli}, {Faisst}, {Ginolfi},
  {B{\'e}thermin}, {Capak}, {Le F{\`e}vre}, {Schaerer}, {Silverman}, {Yan},
  {Lemaux}, {Romano}, {Talia}, {Bardelli}, {Boquien}, {Cimatti},
  {Dessauges-Zavadsky}, {Fudamoto}, {Fujimoto}, {Giavalisco}, {Hathi}, {Ibar},
  {Jones}, {Koekemoer}, {M{\'e}ndez-Hernandez}, {Mancini}, {Oesch}, {Pozzi},
  {Riechers}, {Rodighiero}, {Vergani}, {Zamorani}, \& {Zucca}}]{cassata20}
{Cassata}, P., {Morselli}, L., {Faisst}, A., {et~al.} 2020, \aap, 643, A6

\bibitem[{{Chabrier}(2003)}]{chabrier03}
{Chabrier}, G. 2003, \pasp, 115, 763

\bibitem[{{Cicone} {et~al.}(2014){Cicone}, {Maiolino}, {Sturm},
  {Graci{\'a}-Carpio}, {Feruglio}, {Neri}, {Aalto}, {Davies}, {Fiore},
  {Fischer}, {Garc{\'\i}a-Burillo}, {Gonz{\'a}lez-Alfonso}, {Hailey-Dunsheath},
  {Piconcelli}, \& {Veilleux}}]{cicone14}
{Cicone}, C., {Maiolino}, R., {Sturm}, E., {et~al.} 2014, \aap, 562, A21

\bibitem[{{da Cunha} {et~al.}(2008){da Cunha}, {Charlot}, \& {Elbaz}}]{cuna08}
{da Cunha}, E., {Charlot}, S., \& {Elbaz}, D. 2008, \mnras, 388, 1595

\bibitem[{{Dessauges-Zavadsky} {et~al.}(2020){Dessauges-Zavadsky}, {Ginolfi},
  {Pozzi}, {B{\'e}thermin}, {Le F{\`e}vre}, {Fujimoto}, {Silverman}, {Jones},
  {Vallini}, {Schaerer}, {Faisst}, {Khusanova}, {Fudamoto}, {Cassata},
  {Loiacono}, {Capak}, {Yan}, {Amorin}, {Bardelli}, {Boquien}, {Cimatti},
  {Gruppioni}, {Hathi}, {Ibar}, {Koekemoer}, {Lemaux}, {Narayanan}, {Oesch},
  {Rodighiero}, {Romano}, {Talia}, {Toft}, {Vergani}, {Zamorani}, \&
  {Zucca}}]{mirka20}
{Dessauges-Zavadsky}, M., {Ginolfi}, M., {Pozzi}, F., {et~al.} 2020, \aap, 643,
  A5

\bibitem[{{Dhanda} {et~al.}(2007){Dhanda}, {Baldwin}, {Bentz}, \&
  {Osmer}}]{dhanda07}
{Dhanda}, N., {Baldwin}, J.~A., {Bentz}, M.~C., \& {Osmer}, P.~S. 2007, \apj,
  658, 804

\bibitem[{{Dickinson} {et~al.}(2003){Dickinson}, {Giavalisco}, \& {GOODS
  Team}}]{dickinson03}
{Dickinson}, M., {Giavalisco}, M., \& {GOODS Team}. 2003, in The Mass of
  Galaxies at Low and High Redshift, ed. R.~{Bender} \& A.~{Renzini}, 324

\bibitem[{{Dietrich} {et~al.}(2003){Dietrich}, {Appenzeller}, {Hamann},
  {Heidt}, {J{\"a}ger}, {Vestergaard}, \& {Wagner}}]{dietrich03}
{Dietrich}, M., {Appenzeller}, I., {Hamann}, F., {et~al.} 2003, \aap, 398, 891

\bibitem[{{Dietrich} \& {Wilhelm-Erkens}(2000)}]{dietrich00}
{Dietrich}, M. \& {Wilhelm-Erkens}, U. 2000, \aap, 354, 17

\bibitem[{{Dors} {et~al.}(2014){Dors}, {Cardaci}, {H{\"a}gele}, \&
  {Krabbe}}]{dors14}
{Dors}, O.~L., {Cardaci}, M.~V., {H{\"a}gele}, G.~F., \& {Krabbe}, {\^A}.~C.
  2014, \mnras, 443, 1291

\bibitem[{{Draine} {et~al.}(2014){Draine}, {Aniano}, {Krause}, {Groves},
  {Sandstrom}, {Braun}, {Leroy}, {Klaas}, {Linz}, {Rix}, {Schinnerer},
  {Schmiedeke}, \& {Walter}}]{draine14}
{Draine}, B.~T., {Aniano}, G., {Krause}, O., {et~al.} 2014, \apj, 780, 172

\bibitem[{{Erben} {et~al.}(2005){Erben}, {Schirmer}, {Dietrich}, {Cordes},
  {Haberzettl}, {Hetterscheidt}, {Hildebrandt}, {Schmithuesen}, {Schneider},
  {Simon}, {Deul}, {Hook}, {Kaiser}, {Radovich}, {Benoist}, {Nonino}, {Olsen},
  {Prandoni}, {Wichmann}, {Zaggia}, {Bomans}, {Dettmar}, \&
  {Miralles}}]{erben05}
{Erben}, T., {Schirmer}, M., {Dietrich}, J.~P., {et~al.} 2005, Astronomische
  Nachrichten, 326, 432

\bibitem[{{Faisst} {et~al.}(2020){Faisst}, {Schaerer}, {Lemaux}, {Oesch},
  {Fudamoto}, {Cassata}, {B{\'e}thermin}, {Capak}, {Le F{\`e}vre}, {Silverman},
  {Yan}, {Ginolfi}, {Koekemoer}, {Morselli}, {Amor{\'\i}n}, {Bardelli},
  {Boquien}, {Brammer}, {Cimatti}, {Dessauges-Zavadsky}, {Fujimoto},
  {Gruppioni}, {Hathi}, {Hemmati}, {Ibar}, {Jones}, {Khusanova}, {Loiacono},
  {Pozzi}, {Talia}, {Tasca}, {Riechers}, {Rodighiero}, {Romano}, {Scoville},
  {Toft}, {Vallini}, {Vergani}, {Zamorani}, \& {Zucca}}]{faisst20}
{Faisst}, A.~L., {Schaerer}, D., {Lemaux}, B.~C., {et~al.} 2020, \apjs, 247, 61

\bibitem[{Fan {et~al.}(2006)Fan, Strauss, Richards, Hennawi, Becker, White,
  Diamond-Stanic, Donley, Jiang, Kim, Vestergaard, Young, Gunn, Lupton, Knapp,
  Schneider, Brandt, Bahcall, Barentine, Brinkmann, Brewington, Fukugita,
  Harvanek, Kleinman, Krzesinski, Long, Eric H.~Neilsen, Nitta, Snedden, \&
  Voges}]{fan06}
Fan, X., Strauss, M.~A., Richards, G.~T., {et~al.} 2006, The Astronomical
  Journal, 131, 1203

\bibitem[{{Feltre} {et~al.}(2016){Feltre}, {Charlot}, \& {Gutkin}}]{feltre16}
{Feltre}, A., {Charlot}, S., \& {Gutkin}, J. 2016, \mnras, 456, 3354

\bibitem[{{Ferland} {et~al.}(2013){Ferland}, {Porter}, {van Hoof}, {Williams},
  {Abel}, {Lykins}, {Shaw}, {Henney}, \& {Stancil}}]{ferland13}
{Ferland}, G.~J., {Porter}, R.~L., {van Hoof}, P.~A.~M., {et~al.} 2013, \rmxaa,
  49, 137

\bibitem[{{Ferrarese} \& {Merritt}(2000)}]{ferrarese00}
{Ferrarese}, L. \& {Merritt}, D. 2000, \apjl, 539, L9

\bibitem[{{Feruglio} {et~al.}(2010){Feruglio}, {Maiolino}, {Piconcelli},
  {Menci}, {Aussel}, {Lamastra}, \& {Fiore}}]{feruglio10}
{Feruglio}, C., {Maiolino}, R., {Piconcelli}, E., {et~al.} 2010, \aap, 518,
  L155

\bibitem[{{Fontanot} {et~al.}(2007){Fontanot}, {Cristiani}, {Monaco}, {Nonino},
  {Vanzella}, {Brandt}, {Grazian}, \& {Mao}}]{fontanot07}
{Fontanot}, F., {Cristiani}, S., {Monaco}, P., {et~al.} 2007, \aap, 461, 39

\bibitem[{{Fosbury} {et~al.}(2003){Fosbury}, {Humphrey}, {Villar-Mart{\'\i}n},
  {Rosati}, {Squires}, {Stanford}, {Holden}, \& {Rauch}}]{fosbury03}
{Fosbury}, R.~A.~E., {Humphrey}, A., {Villar-Mart{\'\i}n}, M., {et~al.} 2003,
  in The Mass of Galaxies at Low and High Redshift, ed. R.~{Bender} \&
  A.~{Renzini}, 308

\bibitem[{{Gebhardt} {et~al.}(2000){Gebhardt}, {Bender}, {Bower}, {Dressler},
  {Faber}, {Filippenko}, {Green}, {Grillmair}, {Ho}, {Kormendy}, {Lauer},
  {Magorrian}, {Pinkney}, {Richstone}, \& {Tremaine}}]{gebhardt00}
{Gebhardt}, K., {Bender}, R., {Bower}, G., {et~al.} 2000, \apjl, 539, L13

\bibitem[{{Giacconi} {et~al.}(2002){Giacconi}, {Zirm}, {Wang}, {Rosati},
  {Nonino}, {Tozzi}, {Gilli}, {Mainieri}, {Hasinger}, {Kewley}, {Bergeron},
  {Borgani}, {Gilmozzi}, {Grogin}, {Koekemoer}, {Schreier}, {Zheng}, \&
  {Norman}}]{giacconi02}
{Giacconi}, R., {Zirm}, A., {Wang}, J., {et~al.} 2002, \apjs, 139, 369

\bibitem[{{Giallongo} {et~al.}(2019){Giallongo}, {Grazian}, {Fiore}, {Kodra},
  {Urrutia}, {Castellano}, {Cristiani}, {Dickinson}, {Fontana}, {Menci},
  {Pentericci}, {Boutsia}, {Newman}, \& {Puccetti}}]{giallongo19}
{Giallongo}, E., {Grazian}, A., {Fiore}, F., {et~al.} 2019, \apj, 884, 19

\bibitem[{{Giavalisco} {et~al.}(2004){Giavalisco}, {Ferguson}, {Koekemoer},
  {Dickinson}, {Alexander}, {Bauer}, {Bergeron}, {Biagetti}, {Brandt},
  {Casertano}, {Cesarsky}, {Chatzichristou}, {Conselice}, {Cristiani}, {Da
  Costa}, {Dahlen}, {de Mello}, {Eisenhardt}, {Erben}, {Fall}, {Fassnacht},
  {Fosbury}, {Fruchter}, {Gardner}, {Grogin}, {Hook}, {Hornschemeier}, {Idzi},
  {Jogee}, {Kretchmer}, {Laidler}, {Lee}, {Livio}, {Lucas}, {Madau},
  {Mobasher}, {Moustakas}, {Nonino}, {Padovani}, {Papovich}, {Park},
  {Ravindranath}, {Renzini}, {Richardson}, {Riess}, {Rosati}, {Schirmer},
  {Schreier}, {Somerville}, {Spinrad}, {Stern}, {Stiavelli}, {Strolger},
  {Urry}, {Vandame}, {Williams}, \& {Wolf}}]{giavalisco04}
{Giavalisco}, M., {Ferguson}, H.~C., {Koekemoer}, A.~M., {et~al.} 2004, \apjl,
  600, L93

\bibitem[{{Glikman} {et~al.}(2007){Glikman}, {Djorgovski}, {Stern},
  {Bogosavljevi{\'c}}, \& {Mahabal}}]{glikman07}
{Glikman}, E., {Djorgovski}, S.~G., {Stern}, D., {Bogosavljevi{\'c}}, M., \&
  {Mahabal}, A. 2007, \apjl, 663, L73

\bibitem[{{Grazian} {et~al.}(2020){Grazian}, {Giallongo}, {Fiore}, {Boutsia},
  {Civano}, {Cristiani}, {Cupani}, {Dickinson}, {Fontanot}, {Menci}, \&
  {Romano}}]{grazian20}
{Grazian}, A., {Giallongo}, E., {Fiore}, F., {et~al.} 2020, \apj, 897, 94

\bibitem[{{Grimes} {et~al.}(2007){Grimes}, {Heckman}, {Strickland}, {Dixon},
  {Sembach}, {Overzier}, {Hoopes}, {Aloisi}, \& {Ptak}}]{grimes07}
{Grimes}, J.~P., {Heckman}, T., {Strickland}, D., {et~al.} 2007, \apj, 668, 891

\bibitem[{{Grogin} {et~al.}(2011){Grogin}, {Kocevski}, {Faber}, {Ferguson},
  {Koekemoer}, {Riess}, {Acquaviva}, {Alexander}, {Almaini}, {Ashby}, {Barden},
  {Bell}, {Bournaud}, {Brown}, {Caputi}, {Casertano}, {Cassata}, {Castellano},
  {Challis}, {Chary}, {Cheung}, {Cirasuolo}, {Conselice}, {Roshan Cooray},
  {Croton}, {Daddi}, {Dahlen}, {Dav{\'e}}, {de Mello}, {Dekel}, {Dickinson},
  {Dolch}, {Donley}, {Dunlop}, {Dutton}, {Elbaz}, {Fazio}, {Filippenko},
  {Finkelstein}, {Fontana}, {Gardner}, {Garnavich}, {Gawiser}, {Giavalisco},
  {Grazian}, {Guo}, {Hathi}, {H{\"a}ussler}, {Hopkins}, {Huang}, {Huang},
  {Jha}, {Kartaltepe}, {Kirshner}, {Koo}, {Lai}, {Lee}, {Li}, {Lotz}, {Lucas},
  {Madau}, {McCarthy}, {McGrath}, {McIntosh}, {McLure}, {Mobasher},
  {Moustakas}, {Mozena}, {Nandra}, {Newman}, {Niemi}, {Noeske}, {Papovich},
  {Pentericci}, {Pope}, {Primack}, {Rajan}, {Ravindranath}, {Reddy}, {Renzini},
  {Rix}, {Robaina}, {Rodney}, {Rosario}, {Rosati}, {Salimbeni}, {Scarlata},
  {Siana}, {Simard}, {Smidt}, {Somerville}, {Spinrad}, {Straughn}, {Strolger},
  {Telford}, {Teplitz}, {Trump}, {van der Wel}, {Villforth}, {Wechsler},
  {Weiner}, {Wiklind}, {Wild}, {Wilson}, {Wuyts}, {Yan}, \& {Yun}}]{grogin11A}
{Grogin}, N.~A., {Kocevski}, D.~D., {Faber}, S.~M., {et~al.} 2011, \apjs, 197,
  35

\bibitem[{{Gruppioni} {et~al.}(2013){Gruppioni}, {Pozzi}, {Rodighiero},
  {Delvecchio}, {Berta}, {Pozzetti}, {Zamorani}, {Andreani}, {Cimatti},
  {Ilbert}, {Le Floc'h}, {Lutz}, {Magnelli}, {Marchetti}, {Monaco}, {Nordon},
  {Oliver}, {Popesso}, {Riguccini}, {Roseboom}, {Rosario}, {Sargent},
  {Vaccari}, {Altieri}, {Aussel}, {Bongiovanni}, {Cepa}, {Daddi},
  {Dom{\'\i}nguez-S{\'a}nchez}, {Elbaz}, {F{\"o}rster Schreiber}, {Genzel},
  {Iribarrem}, {Magliocchetti}, {Maiolino}, {Poglitsch}, {P{\'e}rez
  Garc{\'\i}a}, {Sanchez-Portal}, {Sturm}, {Tacconi}, {Valtchanov}, {Amblard},
  {Arumugam}, {Bethermin}, {Bock}, {Boselli}, {Buat}, {Burgarella},
  {Castro-Rodr{\'\i}guez}, {Cava}, {Chanial}, {Clements}, {Conley}, {Cooray},
  {Dowell}, {Dwek}, {Eales}, {Franceschini}, {Glenn}, {Griffin},
  {Hatziminaoglou}, {Ibar}, {Isaak}, {Ivison}, {Lagache}, {Levenson}, {Lu},
  {Madden}, {Maffei}, {Mainetti}, {Nguyen}, {O'Halloran}, {Page}, {Panuzzo},
  {Papageorgiou}, {Pearson}, {P{\'e}rez-Fournon}, {Pohlen}, {Rigopoulou},
  {Rowan-Robinson}, {Schulz}, {Scott}, {Seymour}, {Shupe}, {Smith}, {Stevens},
  {Symeonidis}, {Trichas}, {Tugwell}, {Vigroux}, {Wang}, {Wright}, {Xu},
  {Zemcov}, {Bardelli}, {Carollo}, {Contini}, {Le F{\'e}vre}, {Lilly},
  {Mainieri}, {Renzini}, {Scodeggio}, \& {Zucca}}]{gruppioni13}
{Gruppioni}, C., {Pozzi}, F., {Rodighiero}, G., {et~al.} 2013, \mnras, 432, 23

\bibitem[{{Guo} {et~al.}(2013){Guo}, {Ferguson}, {Giavalisco}, {Barro},
  {Willner}, {Ashby}, {Dahlen}, {Donley}, {Faber}, {Fontana}, {Galametz},
  {Grazian}, {Huang}, {Kocevski}, {Koekemoer}, {Koo}, {McGrath}, {Peth},
  {Salvato}, {Wuyts}, {Castellano}, {Cooray}, {Dickinson}, {Dunlop}, {Fazio},
  {Gardner}, {Gawiser}, {Grogin}, {Hathi}, {Hsu}, {Lee}, {Lucas}, {Mobasher},
  {Nandra}, {Newman}, \& {van der Wel}}]{guo13}
{Guo}, Y., {Ferguson}, H.~C., {Giavalisco}, M., {et~al.} 2013, \apjs, 207, 24

\bibitem[{{Gutkin} {et~al.}(2016){Gutkin}, {Charlot}, \& {Bruzual}}]{gutkin16}
{Gutkin}, J., {Charlot}, S., \& {Bruzual}, G. 2016, \mnras, 462, 1757

\bibitem[{{Hainline} {et~al.}(2011){Hainline}, {Shapley}, {Greene}, \&
  {Steidel}}]{hainline11}
{Hainline}, K.~N., {Shapley}, A.~E., {Greene}, J.~E., \& {Steidel}, C.~C. 2011,
  \apj, 733, 31

\bibitem[{{Harikane} {et~al.}(2022){Harikane}, {Inoue}, {Mawatari},
  {Hashimoto}, {Yamanaka}, {Fudamoto}, {Matsuo}, {Tamura}, {Dayal}, {Yung},
  {Hutter}, {Pacucci}, {Sugahara}, \& {Koekemoer}}]{harikane22}
{Harikane}, Y., {Inoue}, A.~K., {Mawatari}, K., {et~al.} 2022, \apj, 929, 1

\bibitem[{{Hayes} {et~al.}(2016){Hayes}, {Melinder}, {{\"O}stlin}, {Scarlata},
  {Lehnert}, \& {Mannerstr{\"o}m-Jansson}}]{matthew16}
{Hayes}, M., {Melinder}, J., {{\"O}stlin}, G., {et~al.} 2016, \apj, 828, 49

\bibitem[{{Hildebrandt} {et~al.}(2006){Hildebrandt}, {Erben}, {Dietrich},
  {Cordes}, {Haberzettl}, {Hetterscheidt}, {Schirmer}, {Schmithuesen},
  {Schneider}, {Simon}, \& {Trachternach}}]{hildebrandt06}
{Hildebrandt}, H., {Erben}, T., {Dietrich}, J.~P., {et~al.} 2006, \aap, 452,
  1121

\bibitem[{{Hopkins} {et~al.}(2007){Hopkins}, {Lidz}, {Hernquist}, {Coil},
  {Myers}, {Cox}, \& {Spergel}}]{hopkins08}
{Hopkins}, P.~F., {Lidz}, A., {Hernquist}, L., {et~al.} 2007, \apj, 662, 110

\bibitem[{{Hsieh} {et~al.}(2012){Hsieh}, {Wang}, {Hsieh}, {Lin}, {Yan}, {Lim},
  \& {Ho}}]{hsieh12}
{Hsieh}, B.-C., {Wang}, W.-H., {Hsieh}, C.-C., {et~al.} 2012, \apjs, 203, 23

\bibitem[{{Inoue} {et~al.}(2014){Inoue}, {Shimizu}, {Iwata}, \&
  {Tanaka}}]{inoue14}
{Inoue}, A.~K., {Shimizu}, I., {Iwata}, I., \& {Tanaka}, M. 2014, \mnras, 442,
  1805

\bibitem[{{Jaskot} {et~al.}(2017){Jaskot}, {Oey}, {Scarlata}, \&
  {Dowd}}]{jaskot17}
{Jaskot}, A.~E., {Oey}, M.~S., {Scarlata}, C., \& {Dowd}, T. 2017, \apjl, 851,
  L9

\bibitem[{{Koekemoer} {et~al.}(2011){Koekemoer}, {Faber}, {Ferguson}, {Grogin},
  {Kocevski}, {Koo}, {Lai}, {Lotz}, {Lucas}, {McGrath}, {Ogaz}, {Rajan},
  {Riess}, {Rodney}, {Strolger}, {Casertano}, {Castellano}, {Dahlen},
  {Dickinson}, {Dolch}, {Fontana}, {Giavalisco}, {Grazian}, {Guo}, {Hathi},
  {Huang}, {van der Wel}, {Yan}, {Acquaviva}, {Alexander}, {Almaini}, {Ashby},
  {Barden}, {Bell}, {Bournaud}, {Brown}, {Caputi}, {Cassata}, {Challis},
  {Chary}, {Cheung}, {Cirasuolo}, {Conselice}, {Roshan Cooray}, {Croton},
  {Daddi}, {Dav{\'e}}, {de Mello}, {de Ravel}, {Dekel}, {Donley}, {Dunlop},
  {Dutton}, {Elbaz}, {Fazio}, {Filippenko}, {Finkelstein}, {Frazer}, {Gardner},
  {Garnavich}, {Gawiser}, {Gruetzbauch}, {Hartley}, {H{\"a}ussler},
  {Herrington}, {Hopkins}, {Huang}, {Jha}, {Johnson}, {Kartaltepe},
  {Khostovan}, {Kirshner}, {Lani}, {Lee}, {Li}, {Madau}, {McCarthy},
  {McIntosh}, {McLure}, {McPartland}, {Mobasher}, {Moreira}, {Mortlock},
  {Moustakas}, {Mozena}, {Nandra}, {Newman}, {Nielsen}, {Niemi}, {Noeske},
  {Papovich}, {Pentericci}, {Pope}, {Primack}, {Ravindranath}, {Reddy},
  {Renzini}, {Rix}, {Robaina}, {Rosario}, {Rosati}, {Salimbeni}, {Scarlata},
  {Siana}, {Simard}, {Smidt}, {Snyder}, {Somerville}, {Spinrad}, {Straughn},
  {Telford}, {Teplitz}, {Trump}, {Vargas}, {Villforth}, {Wagner}, {Wandro},
  {Wechsler}, {Weiner}, {Wiklind}, {Wild}, {Wilson}, {Wuyts}, \&
  {Yun}}]{koekemoer11}
{Koekemoer}, A.~M., {Faber}, S.~M., {Ferguson}, H.~C., {et~al.} 2011, \apjs,
  197, 36

\bibitem[{{Koptelova} {et~al.}(2019){Koptelova}, {Hwang}, {Malkan}, \&
  {Yu}}]{koptelova19}
{Koptelova}, E., {Hwang}, C.-Y., {Malkan}, M.~A., \& {Yu}, P.-C. 2019, \apj,
  882, 144

\bibitem[{{Kormendy} \& {Ho}(2013)}]{kormendy13}
{Kormendy}, J. \& {Ho}, L.~C. 2013, \araa, 51, 511

\bibitem[{{Kormendy} \& {Richstone}(1995)}]{kormendy95}
{Kormendy}, J. \& {Richstone}, D. 1995, \araa, 33, 581

\bibitem[{{Laporte} {et~al.}(2021){Laporte}, {Meyer}, {Ellis}, {Robertson},
  {Chisholm}, \& {Roberts-Borsani}}]{laporte21}
{Laporte}, N., {Meyer}, R.~A., {Ellis}, R.~S., {et~al.} 2021, \mnras, 505, 3336

\bibitem[{{Le F{\`e}vre} {et~al.}(2020){Le F{\`e}vre}, {B{\'e}thermin},
  {Faisst}, {Jones}, {Capak}, {Cassata}, {Silverman}, {Schaerer}, {Yan},
  {Amorin}, {Bardelli}, {Boquien}, {Cimatti}, {Dessauges-Zavadsky},
  {Giavalisco}, {Hathi}, {Fudamoto}, {Fujimoto}, {Ginolfi}, {Gruppioni},
  {Hemmati}, {Ibar}, {Koekemoer}, {Khusanova}, {Lagache}, {Lemaux}, {Loiacono},
  {Maiolino}, {Mancini}, {Narayanan}, {Morselli}, {M{\'e}ndez-Hern{\`a}ndez},
  {Oesch}, {Pozzi}, {Romano}, {Riechers}, {Scoville}, {Talia}, {Tasca},
  {Thomas}, {Toft}, {Vallini}, {Vergani}, {Walter}, {Zamorani}, \&
  {Zucca}}]{lefevre20}
{Le F{\`e}vre}, O., {B{\'e}thermin}, M., {Faisst}, A., {et~al.} 2020, \aap,
  643, A1

\bibitem[{{Le F{\`e}vre} {et~al.}(2019){Le F{\`e}vre}, {Lemaux}, {Nakajima},
  {Schaerer}, {Talia}, {Zamorani}, {Cassata}, {Garilli}, {Maccagni},
  {Pentericci}, {Tasca}, {Zucca}, {Amorin}, {Bardelli}, {Cimatti},
  {Giavalisco}, {Guaita}, {Hathi}, {Marchi}, {Vanzella}, {Vergani}, \&
  {Dunlop}}]{lefevre19}
{Le F{\`e}vre}, O., {Lemaux}, B.~C., {Nakajima}, K., {et~al.} 2019, \aap, 625,
  A51

\bibitem[{{Leitherer} {et~al.}(1999){Leitherer}, {Schaerer}, {Goldader},
  {Delgado}, {Robert}, {Kune}, {de Mello}, {Devost}, \&
  {Heckman}}]{leitherer99}
{Leitherer}, C., {Schaerer}, D., {Goldader}, J.~D., {et~al.} 1999, \apjs, 123,
  3

\bibitem[{{Lin} {et~al.}(2022){Lin}, {Scarlata}, {Hayes}, {Feltre}, {Charlot},
  {Bongiorno}, {V{\"a}is{\"a}nen}, \& {Mogotsi}}]{lin22}
{Lin}, Y.-H., {Scarlata}, C., {Hayes}, M., {et~al.} 2022, \mnras, 509, 489

\bibitem[{{Lusso} {et~al.}(2012){Lusso}, {Comastri}, {Simmons}, {Mignoli},
  {Zamorani}, {Vignali}, {Brusa}, {Shankar}, {Lutz}, {Trump}, {Maiolino},
  {Gilli}, {Bolzonella}, {Puccetti}, {Salvato}, {Impey}, {Civano}, {Elvis},
  {Mainieri}, {Silverman}, {Koekemoer}, {Bongiorno}, {Merloni}, {Berta}, {Le
  Floc'h}, {Magnelli}, {Pozzi}, \& {Riguccini}}]{lusso12}
{Lusso}, E., {Comastri}, A., {Simmons}, B.~D., {et~al.} 2012, \mnras, 425, 623

\bibitem[{{Magorrian} {et~al.}(1998){Magorrian}, {Tremaine}, {Richstone},
  {Bender}, {Bower}, {Dressler}, {Faber}, {Gebhardt}, {Green}, {Grillmair},
  {Kormendy}, \& {Lauer}}]{magorrian98}
{Magorrian}, J., {Tremaine}, S., {Richstone}, D., {et~al.} 1998, \aj, 115, 2285

\bibitem[{{Marques-Chaves} {et~al.}(2021){Marques-Chaves}, {Schaerer},
  {{\'A}lvarez-M{\'a}rquez}, {Colina}, {Dessauges-Zavadsky},
  {P{\'e}rez-Fournon}, {Saldana-Lopez}, \& {Verhamme}}]{rui21}
{Marques-Chaves}, R., {Schaerer}, D., {{\'A}lvarez-M{\'a}rquez}, J., {et~al.}
  2021, \mnras, 507, 524

\bibitem[{{Matthee} {et~al.}(2022){Matthee}, {Feltre}, {Maseda}, {Nanayakkara},
  {Boogaard}, {Bacon}, {Verhamme}, {Leclercq}, {Kusakabe}, {Urrutia}, \&
  {Wisotzki}}]{matthee22}
{Matthee}, J., {Feltre}, A., {Maseda}, M., {et~al.} 2022, \aap, 660, A10

\bibitem[{{McGreer} {et~al.}(2018){McGreer}, {Cl{\'e}ment}, {Mainali}, {Stark},
  {Gronke}, {Dijkstra}, {Fan}, {Bian}, {Frye}, {Jiang}, {Kneib}, {Limousin}, \&
  {Walth}}]{mcgreer18}
{McGreer}, I.~D., {Cl{\'e}ment}, B., {Mainali}, R., {et~al.} 2018, \mnras, 479,
  435

\bibitem[{{McKernan} {et~al.}(2007){McKernan}, {Yaqoob}, \&
  {Reynolds}}]{mckernan07}
{McKernan}, B., {Yaqoob}, T., \& {Reynolds}, C.~S. 2007, \mnras, 379, 1359

\bibitem[{{McLure} {et~al.}(2018){McLure}, {Pentericci}, {Cimatti}, {Dunlop},
  {Elbaz}, {Fontana}, {Nandra}, {Amorin}, {Bolzonella}, {Bongiorno}, {Carnall},
  {Castellano}, {Cirasuolo}, {Cucciati}, {Cullen}, {De Barros}, {Finkelstein},
  {Fontanot}, {Franzetti}, {Fumana}, {Gargiulo}, {Garilli}, {Guaita},
  {Hartley}, {Iovino}, {Jarvis}, {Juneau}, {Karman}, {Maccagni}, {Marchi},
  {M{\'a}rmol-Queralt{\'o}}, {Pompei}, {Pozzetti}, {Scodeggio}, {Sommariva},
  {Talia}, {Almaini}, {Balestra}, {Bardelli}, {Bell}, {Bourne}, {Bowler},
  {Brusa}, {Buitrago}, {Caputi}, {Cassata}, {Charlot}, {Citro}, {Cresci},
  {Cristiani}, {Curtis-Lake}, {Dickinson}, {Fazio}, {Ferguson}, {Fiore},
  {Franco}, {Fynbo}, {Galametz}, {Georgakakis}, {Giavalisco}, {Grazian},
  {Hathi}, {Jung}, {Kim}, {Koekemoer}, {Khusanova}, {Le F{\`e}vre}, {Lotz},
  {Mannucci}, {Maltby}, {Matsuoka}, {McLeod}, {Mendez-Hernandez},
  {Mendez-Abreu}, {Mignoli}, {Moresco}, {Mortlock}, {Nonino}, {Pannella},
  {Papovich}, {Popesso}, {Rosario}, {Salvato}, {Santini}, {Schaerer},
  {Schreiber}, {Stark}, {Tasca}, {Thomas}, {Treu}, {Vanzella}, {Wild},
  {Williams}, {Zamorani}, \& {Zucca}}]{mclure18}
{McLure}, R.~J., {Pentericci}, L., {Cimatti}, A., {et~al.} 2018, \mnras, 479,
  25

\bibitem[{{Morisset} {et~al.}(2015){Morisset}, {Delgado-Inglada}, \&
  {Flores-Fajardo}}]{morisset15}
{Morisset}, C., {Delgado-Inglada}, G., \& {Flores-Fajardo}, N. 2015, \rmxaa,
  51, 103

\bibitem[{{Noll} {et~al.}(2009){Noll}, {Burgarella}, {Giovannoli}, {Buat},
  {Marcillac}, \& {Mu{\~n}oz-Mateos}}]{noll09}
{Noll}, S., {Burgarella}, D., {Giovannoli}, E., {et~al.} 2009, \aap, 507, 1793

\bibitem[{{Nonino} {et~al.}(2009){Nonino}, {Dickinson}, {Rosati}, {Grazian},
  {Reddy}, {Cristiani}, {Giavalisco}, {Kuntschner}, {Vanzella}, {Daddi},
  {Fosbury}, \& {Cesarsky}}]{nonino09}
{Nonino}, M., {Dickinson}, M., {Rosati}, P., {et~al.} 2009, \apjs, 183, 244

\bibitem[{{Orr} {et~al.}(2019){Orr}, {Hayward}, \& {Hopkins}}]{orr19}
{Orr}, M.~E., {Hayward}, C.~C., \& {Hopkins}, P.~F. 2019, \mnras, 486, 4724

\bibitem[{{Otte} {et~al.}(2003){Otte}, {Murphy}, {Howk}, {Wang}, {Oegerle}, \&
  {Sembach}}]{otte03}
{Otte}, B., {Murphy}, E.~M., {Howk}, J.~C., {et~al.} 2003, \apj, 591, 821

\bibitem[{{Patr{\'\i}cio} {et~al.}(2016){Patr{\'\i}cio}, {Richard}, {Verhamme},
  {Wisotzki}, {Brinchmann}, {Turner}, {Christensen}, {Weilbacher}, {Blaizot},
  {Bacon}, {Contini}, {Lagattuta}, {Cantalupo}, {Cl{\'e}ment}, \&
  {Soucail}}]{patricio16}
{Patr{\'\i}cio}, V., {Richard}, J., {Verhamme}, A., {et~al.} 2016, \mnras, 456,
  4191

\bibitem[{{Prinja} \& {Howarth}(1986)}]{prinja86}
{Prinja}, R.~K. \& {Howarth}, I.~D. 1986, \apjs, 61, 357

\bibitem[{{Raiter} {et~al.}(2010){Raiter}, {Fosbury}, \&
  {Teimoorinia}}]{raiter10}
{Raiter}, A., {Fosbury}, R.~A.~E., \& {Teimoorinia}, H. 2010, \aap, 510, A109

\bibitem[{{Retzlaff} {et~al.}(2010){Retzlaff}, {Rosati}, {Dickinson},
  {Vandame}, {Rit{\'e}}, {Nonino}, {Cesarsky}, \& {GOODS Team}}]{retzlaff10}
{Retzlaff}, J., {Rosati}, P., {Dickinson}, M., {et~al.} 2010, \aap, 511, A50

\bibitem[{{Rupke} {et~al.}(2005){Rupke}, {Veilleux}, \& {Sanders}}]{rupke05}
{Rupke}, D.~S., {Veilleux}, S., \& {Sanders}, D.~B. 2005, \apjs, 160, 115

\bibitem[{{Schaerer} {et~al.}(2020){Schaerer}, {Ginolfi}, {B{\'e}thermin},
  {Fudamoto}, {Oesch}, {Le F{\`e}vre}, {Faisst}, {Capak}, {Cassata},
  {Silverman}, {Yan}, {Jones}, {Amorin}, {Bardelli}, {Boquien}, {Cimatti},
  {Dessauges-Zavadsky}, {Giavalisco}, {Hathi}, {Fujimoto}, {Ibar}, {Koekemoer},
  {Lagache}, {Lemaux}, {Loiacono}, {Maiolino}, {Narayanan}, {Morselli},
  {M{\'e}ndez-Hern{\`a}ndez}, {Pozzi}, {Riechers}, {Talia}, {Toft}, {Vallini},
  {Vergani}, {Zamorani}, \& {Zucca}}]{schaerer20}
{Schaerer}, D., {Ginolfi}, M., {B{\'e}thermin}, M., {et~al.} 2020, \aap, 643,
  A3

\bibitem[{{Schartmann} {et~al.}(2005){Schartmann}, {Meisenheimer}, {Camenzind},
  {Wolf}, \& {Henning}}]{schartmann05}
{Schartmann}, M., {Meisenheimer}, K., {Camenzind}, M., {Wolf}, S., \&
  {Henning}, T. 2005, \aap, 437, 861

\bibitem[{{Speagle} {et~al.}(2014){Speagle}, {Steinhardt}, {Capak}, \&
  {Silverman}}]{speagle14}
{Speagle}, J.~S., {Steinhardt}, C.~L., {Capak}, P.~L., \& {Silverman}, J.~D.
  2014, \apjs, 214, 15

\bibitem[{{Spergel} {et~al.}(2003){Spergel}, {Verde}, {Peiris}, {Komatsu},
  {Nolta}, {Bennett}, {Halpern}, {Hinshaw}, {Jarosik}, {Kogut}, {Limon},
  {Meyer}, {Page}, {Tucker}, {Weiland}, {Wollack}, \& {Wright}}]{spergel03}
{Spergel}, D.~N., {Verde}, L., {Peiris}, H.~V., {et~al.} 2003, \apjs, 148, 175

\bibitem[{{Stalevski} {et~al.}(2012){Stalevski}, {Fritz}, {Baes}, {Nakos}, \&
  {Popovic}}]{stalevski12}
{Stalevski}, M., {Fritz}, J., {Baes}, M., {Nakos}, T., \& {Popovic}, L.~C.
  2012, Publications de l'Observatoire Astronomique de Beograd, 91, 235

\bibitem[{{Stalevski} {et~al.}(2016){Stalevski}, {Ricci}, {Ueda}, {Lira},
  {Fritz}, \& {Baes}}]{stalevski16}
{Stalevski}, M., {Ricci}, C., {Ueda}, Y., {et~al.} 2016, \mnras, 458, 2288

\bibitem[{{Stanway} \& {Eldridge}(2018)}]{stanway18}
{Stanway}, E.~R. \& {Eldridge}, J.~J. 2018, \mnras, 479, 75

\bibitem[{{Stark} {et~al.}(2014){Stark}, {Richard}, {Siana}, {Charlot},
  {Freeman}, {Gutkin}, {Wofford}, {Robertson}, {Amanullah}, {Watson}, \&
  {Milvang-Jensen}}]{stark14}
{Stark}, D.~P., {Richard}, J., {Siana}, B., {et~al.} 2014, \mnras, 445, 3200

\bibitem[{{Tacchella} {et~al.}(2016){Tacchella}, {Dekel}, {Carollo},
  {Ceverino}, {DeGraf}, {Lapiner}, {Mandelker}, \& {Primack}}]{tacchella16}
{Tacchella}, S., {Dekel}, A., {Carollo}, C.~M., {et~al.} 2016, \mnras, 458, 242

\bibitem[{{Tang} {et~al.}(2021){Tang}, {Stark}, {Chevallard}, {Charlot},
  {Endsley}, \& {Congiu}}]{tang21}
{Tang}, M., {Stark}, D.~P., {Chevallard}, J., {et~al.} 2021, \mnras, 501, 3238

\bibitem[{{van Dokkum} {et~al.}(2013){van Dokkum}, {Brammer}, {Momcheva},
  {Skelton}, \& {Whitaker}}]{vandokkum13}
{van Dokkum}, P., {Brammer}, G., {Momcheva}, I., {Skelton}, R.~E., \&
  {Whitaker}, K.~E. 2013, arXiv e-prints, arXiv:1305.2140

\bibitem[{{Vanzella} {et~al.}(2006){Vanzella}, {Cristiani}, {Dickinson},
  {Kuntschner}, {Nonino}, {Rettura}, {Rosati}, {Vernet}, {Cesarsky},
  {Ferguson}, {Fosbury}, {Giavalisco}, {Grazian}, {Haase}, {Moustakas},
  {Popesso}, {Renzini}, {Stern}, \& {GOODS Team}}]{vanzella06}
{Vanzella}, E., {Cristiani}, S., {Dickinson}, M., {et~al.} 2006, \aap, 454, 423

\bibitem[{{Vanzella} {et~al.}(2010){Vanzella}, {Grazian}, {Hayes},
  {Pentericci}, {Schaerer}, {Dickinson}, {Cristiani}, {Giavalisco}, {Verhamme},
  {Nonino}, \& {Rosati}}]{vanzella10}
{Vanzella}, E., {Grazian}, A., {Hayes}, M., {et~al.} 2010, \aap, 513, A20

\bibitem[{{Vanzella} {et~al.}(2018){Vanzella}, {Nonino}, {Cupani},
  {Castellano}, {Sani}, {Mignoli}, {Calura}, {Meneghetti}, {Gilli}, {Comastri},
  {Mercurio}, {Caminha}, {Caputi}, {Rosati}, {Grillo}, {Cristiani}, {Balestra},
  {Fontana}, \& {Giavalisco}}]{vanzella18}
{Vanzella}, E., {Nonino}, M., {Cupani}, G., {et~al.} 2018, \mnras, 476, L15

\bibitem[{{Vietri} {et~al.}(2022){Vietri}, {Misawa}, {Piconcelli}, {Franzetti},
  {Luminari}, {Travascio}, {Bischetti}, {Bisogni}, {Bongiorno}, {Bruni},
  {Feruglio}, {Giunta}, {Nicastro}, {Saccheo}, {Testa}, {Tombesi}, {Vignali},
  {Zappacosta}, \& {Fiore}}]{vietri22}
{Vietri}, G., {Misawa}, T., {Piconcelli}, E., {et~al.} 2022, arXiv e-prints,
  arXiv:2205.06832

\bibitem[{{Vito} {et~al.}(2022){Vito}, {Mignoli}, {Gilli}, {Brandt}, {Shemmer},
  {Bauer}, {Bisogni}, {Luo}, {Marchesi}, {Nanni}, {Zamorani}, {Comastri},
  {Cusano}, {Gallerani}, {Vignali}, \& {Lanzuisi}}]{vito22}
{Vito}, F., {Mignoli}, M., {Gilli}, R., {et~al.} 2022, \aap, 663, A159

\bibitem[{{Weymann} {et~al.}(1991){Weymann}, {Morris}, {Foltz}, \&
  {Hewett}}]{weymann91}
{Weymann}, R.~J., {Morris}, S.~L., {Foltz}, C.~B., \& {Hewett}, P.~C. 1991,
  \apj, 373, 23

\bibitem[{{Wiklind} {et~al.}(2008){Wiklind}, {Dickinson}, {Ferguson},
  {Giavalisco}, {Mobasher}, {Grogin}, \& {Panagia}}]{wiklind08}
{Wiklind}, T., {Dickinson}, M., {Ferguson}, H.~C., {et~al.} 2008, \apj, 676,
  781

\bibitem[{{Wuyts} {et~al.}(2008){Wuyts}, {Labb{\'e}}, {F{\"o}rster Schreiber},
  {Franx}, {Rudnick}, {Brammer}, \& {van Dokkum}}]{wuyts08}
{Wuyts}, S., {Labb{\'e}}, I., {F{\"o}rster Schreiber}, N.~M., {et~al.} 2008,
  \apj, 682, 985

\bibitem[{{Yang} {et~al.}(2022){Yang}, {Boquien}, {Brandt}, {Buat},
  {Burgarella}, {Ciesla}, {Lehmer}, {Ma{\l}ek}, {Mountrichas}, {Papovich},
  {Pons}, {Stalevski}, {Theul{\'e}}, \& {Zhu}}]{yang22}
{Yang}, G., {Boquien}, M., {Brandt}, W.~N., {et~al.} 2022, \apj, 927, 192

\bibitem[{{Yang} {et~al.}(2020){Yang}, {Boquien}, {Buat}, {Burgarella},
  {Ciesla}, {Duras}, {Stalevski}, {Brandt}, \& {Papovich}}]{yang20}
{Yang}, G., {Boquien}, M., {Buat}, V., {et~al.} 2020, \mnras, 491, 740

\bibitem[{{Zana} {et~al.}(2022){Zana}, {Gallerani}, {Carniani}, {Vito},
  {Ferrara}, {Lupi}, {Di Mascia}, \& {Barai}}]{zana22}
{Zana}, T., {Gallerani}, S., {Carniani}, S., {et~al.} 2022, \mnras, 513, 2118

\end{thebibliography}

\clearpage
\onecolumn
\begin{appendix}

\section{\ovi, \nv, \niv\ literature sources}\label{sec:app_sources}

\begin{table*}[h!]
\caption{Compilation of literature sources with \ovi, \nv, or \niv\ emission lines. Each row refers to a single source, a collection of sources, or a stacked spectrum. In the first case, the \textit{ID} columns refer to the source name, in the second case, to the number and type of sources in the collection, and in the third case, to the number of sources included in the stack. \ovi, \nv, and \niv\ report the detection (\checkmark) or the non-detection ($\times$) of the respective emission line. No symbol indicates that the line falls out of the spectral range. In case of a collection of sources, the percentage refers to the fraction of sources with detection in the reported emission lines. Finally, the \textit{p} indicates that the line exhibits a P-Cygni profile.
}
\label{tab:literature_sources}
\centering
\begin{tabular}{ccccccc}
\hline\hline
& ID & $z$ & \ovi & \nv & \niv & Ref \\
\hline
 & GS+14 & 5.55 & \checkmark & \checkmark p & \checkmark & This work \\
\hline 
\multirow{18}{*}{AGN}  & Q0353-383 & 1.96 & \checkmark & \checkmark & \checkmark & \citet{baldwin03}\\
 & DLS1053-0528 & 4.02 & & $\times$ & \checkmark & \citet{glikman07}\\
 & NDWFS1433 & 3.88 & & \checkmark & \checkmark & \citet{glikman07}\\
 & UDS24561 & 3.21 & \checkmark & $\times$ & $\times$ & \citet{tang21}\\
 & 33 NL AGN & $2-3$ & \checkmark & \checkmark & \checkmark & \citet{hainline11} \\
 & 12 Seyfert 2 & $0-4$ & & \checkmark $60\%$ & & \citet{dors14}\\
 & 59 radio-galaxies & $0-4$ & & \checkmark $25\%$ & & \citet{dors14}\\
 & 10 QSO2 & $0-4$ & & \checkmark $70\%$ & & \citet{dors14}\\
 & S82-20 & 3.08 & \checkmark & \checkmark & & \citet{lin22}\\
 & J1254+0241 & 1.8 & & \checkmark & \checkmark & \citet{dhanda07}\\
 & J1546-5253 & 2.0 & & \checkmark & \checkmark & \citet{dhanda07}\\
 & J1553+0056 & 2.63 & & \checkmark p & $\times$ & \citet{appenzeller15} \\
 & 16 QSO & $2.4-3.8$ &  \checkmark$26\%$ & \checkmark & & \citet{dietrich00} \\
 & 11 QSO & $3.9-5.0$ &  \checkmark$73\%$ & \checkmark & \checkmark$73\%$ & \citet{dietrich03} \\
 & PSO J006+39 & 6.61 & & \checkmark & $\times$ & \citet{koptelova19} \\
 & J1512+119 & 2.11 & \checkmark & $\times$ &  $\times$ & \citet{borguet12} \\
 & J000239+255034 & 5.80 & & \checkmark & & \citet{fan06} \\
 & 124 QSO stack & $5.6-6.7$ & & \checkmark & & \citet{banados16} \\
 & 13 AGN stack & $2-3.8$ & & \checkmark & \checkmark & \citep{lefevre19} \\
\hline
\multirow{9}{*}{galaxy} & Lynx Arc & 3.36 & $\times$ & $\times$ & \checkmark & \citet{fosbury03} \\
 & 31 stack EW$_{\cii}>20$ &$2-3.8$ & & \checkmark & \checkmark & \citet{lefevre19}\\
 & 120 stack $ 10<\rm{EW}_{\cii}<20$ &$2-3.8$ & & \checkmark & \checkmark & \citet{lefevre19}\\
 & J141445+544631 & 5.42 & $\times$ & $\times$ & \checkmark & \citet{mcgreer18} \\
 & J160810+352809 & 0.03 & & \checkmark p & & \citet{jaskot17} \\
 & Ion3 & 4.0 & $\times$ & \checkmark p & $\times$ & \citet{vanzella18} \\
 & 17 low mass galaxies stack & $1.4-2.9$ & &  \checkmark  & \checkmark $6\%$ & \citet{stark14} \\
 & L$^*$ lensed galaxy & 3.5 & & \checkmark & \checkmark & \citet{patricio16} \\
 & J0121+0025 & 3.24 & \checkmark p & \checkmark p & $\times$ & \citet{rui21} \\
\hline
\end{tabular}
\end{table*}

\newpage
\section{Photometry}\label{sec:app_filters}
We summarise here the ground- and space- based photometric data used in this work for the SED-fitting (see §\ref{sec:sedfitting}). The adopted filters, as well as their wavelengths and references to the measurements, are reported in Table~\ref{tab:photom_band}. The photometry of all ALPINE sources has been collected and calibrated by \citet{faisst20}. It comes primarily from the 3D-\textit{HST} catalog and has been corrected for Galactic extinction, PSF size and other biases. Some additional data, not present in the 3D-\textit{HST} catalog, comes from various observation programs in the Extended Chandra Deep Field South \citet[ECDFS,]{giacconi02} and has been measured and calibrated by \citet{faisst20}. GS-14 has not been detected by MPG-ESO/WFI in band $U38$, $b$, and $v$, in \textit{HST}/ACS $F435W$ band, by Subaru/Suprime-Cam $IA445$, $IA505$, $IA527$, $IA550$, $IA574$, $IA598$, $IA624$, and $IA738$ filters, and by \textit{Spitzer/MIPS} at $24\,\mu$m. Regarding the X-ray bands, GS-14 is undetected in the ultradeep 7 Ms X-ray image by \textit{Chandra} with a flux limit of $10^{-17}\,\rm{erg\,cm^{-2}\,s^{-1}}$ in the observed-frame $0.5-2.0\,$keV band \citep{giallongo19}. We refer to \citet{faisst20} for further details.
\begin{table}[h!]
\caption{Photometric bands used in this work. For details about the data, extraction, and calibration, we refer to \citet{faisst20}.  References: (1) \citet{hildebrandt06, erben05}, (2) \citet{wuyts08,retzlaff10}, (3) \citet{hsieh12}, (4) \citet{cardamone10}, (5) \citet{giavalisco04}, (6) \citet{koekemoer11}, (7) \citet{grogin11A,koekemoer11}, (8) \citet{brammer12,vandokkum13}, (9) \citet{ashby13,guo13}, (10) \citet{dickinson03}.}
\label{tab:photom_band}
\centering
\begin{tabular}{cccc}
\hline \hline
Observatory & Filter & Central & Ref.    \\
/Instrument & & $\lambda$ [\AA] & \\
\hline\\[-9pt]
MPG-ESO/WFI & \textit{I} & 3633.3 & 1 \\[3pt]
VLT/ISAAC	& $J^v$ 	& 	12492.2 	& 	2 \\
&$H^v$ & 16519.9 & 2 \\
 & $K_s^v$ & 21638.3 & 2 \\[3pt]
CFHT/WIRCam & $J^w$ & 12544.6 & 3 \\
 & $K_s^w$ & 21590.4 & 3 \\[3pt]
Subaru/Suprime-Cam & $IA856$ & 8566.0 & 4 \\[3pt]
 \textit{HST}/ACS & $F814W$ & 8058.2 & 5 \\
 & $F850LP$ & 9181.2 & 6 \\[3pt]
 \textit{HST}/WFC3& $F125W$ & 12516.3 & 7 \\
 & $F140W$ & 13969.4 & 8 \\
 & $F160W$ & 15391.1 & 7\\[3pt]
\textit{Spitzer}/IRAC	&	ch1	&	35634.3	&	9	\\
	&	ch2	&	45110.1	&	9	\\
	&	ch3	&	57593.4	&	10	\\
	&	ch4	&	79594.9	&	10	\\ 	
\hline
\end{tabular}
\end{table}

\newpage
\section{\texttt{X-CIGALE} SED-fitting parameter space}\label{sec:app_sed}
\begin{table}[h!]
\caption{Parameter space for the \texttt{X-CIGALE} SED fitting. For each parameter, \textit{N$_{\rm{sample}}$} values are simulated in the \textit{values} range. As the best-fits were obtained with a double exponential SFH (\textit{sfh2exp}), we report only the parameter space for this SFH, while the other investigated SFHs are indicated in \textit{SFH}: the delayed SFH (\textit{sfhdelayed}), the delayed SFH with burst (\textit{sfhdelayedbq}), and the periodic SFH (\textit{sfhperiodic}).}
\label{tab:param_space_sed}
\centering
\begin{tabular}{cccc}
 & N$_{\rm{sample}}$ & values & description \\
\hline \hline
SFH & & sfh2exp; sfhdelayed; & double exponential; delayed SFH with optional exponential burst;\\
& & sfhdelayedbq; sfhperiodic & delayed SFH with optional constant burst/quench; periodic SFH.\\
$\tau_{\rm{main}}$ & 3 & $50,500,1000$ & e-folding time of the main stellar population model in Myr.\\
age$_{\rm{main}}$& 6 & $100,250,500,600,700,1000$& Age of the main stellar population in the galaxy in Myr.\\
$\tau_{\rm{burst}}$& 3 & $50,100,500$& e-folding time of the late starburst population model in Myr.\\
age$_{\rm{burst}}$ & 4 & 5,10,20,50 & Age of the late burst in Myr.\\
f$_{\rm{burst}}$& 4 & $0,0.001,0.01,0.1$&Mass fraction of the late burst population.\\
\hline
IMF& & Chabrier & Initial Mass Function\\
$Z/\rm{Z_{\odot}}$ & 4 & $0.03, 0.3, 1.3,3.2$&Metallicity\\
\hline
$\log \rm{U}_{S}$ & 2 & $-2,-1$&Nebular component: ionization parameter\\
Z$_{\rm{gas}}$& 2 & $0.0004,0.004$&Nebular component: gas metallicity\\
\hline
dust attenaution& & Charlot\&Fall 2000\\
A$_{\rm{V,ISM}}$& 3 & $0.3,1.7,3.3$&V-band attenuation in the interstellar medium.\\
$\mu$& 3 & $0.3,0.5,1.0$&A$_{\rm{V,ISM}}$ / (A$_{\rm{V,BC}}$+A$_{\rm{V,ISM}}$)\\
\hline
Dust emission& & Draine+2014&\\
q$_{\rm{PAH}}$& 3 & $0.47,2.5,3.9$&Mass fraction of PAH.\\
U$_{\rm{min}}$& 4 & $5,10,25,40$&Minimum radiation field.\\
$\alpha$& 1 & $2.0$&Powerlaw slope dU/dM $\propto$U$^\alpha$.\\
$\gamma$& 2 & $0.02,0.1$&Fraction illuminated from U$_{\rm{min}}$ to U$_{\rm{max}}$.\\
\hline
AGN& & Skirtor16\\
$\tau_{9.7\mu\rm{m}}$& 5 & $3,5,7,9,11$& Average edge-on optical depth at 9.7$\,\mu$m.\\
\textit{p} & 1 & 1.0 & Power-law index of radial gradient of dust density.\\
\textit{q} & 1 & 1.0 & Power-law index of angular gradient of dust density.\\ 
\textit{oa} & 1 & 40\textdegree & Torus half-opening angle.\\
R$_{\rm{ratio}}$& 1 & $20$&Ratio of the maximum to minimum radii of the dust torus.\\
M$_{\rm{cl}}$ & 1 & 0.97 & Mass fraction of dust inside clumps.\\
$i$ & 6 & 0\textdegree, 20\textdegree, 40\textdegree, 60\textdegree, 70\textdegree, 90\textdegree & Viewing angle (w.r.t. the AGN axis).\\
disk type & 1 & \citet{schartmann05} & Disk spectrum.\\
$\delta$ & 1 & -0.36 & Power-law index modifying the optical slope of the disk.\\
f$_{\rm{AGN}}$& 4 & $0.001,0.01,0.05,0.1$& $8-1000\,\mu$m AGN fraction.\\
law & 1 & SMC & Extinction law of polar dust.\\
E(B-V) & 1 & 0.03 & E(B-V) for the extinction in the polar direction in magnitudes.\\
T$_{\rm{dust}}^{\rm{polar}}$  & 1 & $100\,$K & Temperature of the polar dust.\\
$\epsilon$ & 1 & 1.6 & Emissivity index of the polar dust.\\
\hline
\end{tabular}
\end{table}

\end{appendix}
\end{document}